\documentclass[preprint]{aastex}

\shorttitle{Observations of Subaru/HDS}
\shortauthors{Honda et al.}

\begin{document}

\title{Spectroscopic Studies of Extremely Metal-Poor Stars with Subaru/HDS: \\
I. Observational Data. \altaffilmark{1}}

\author{Satoshi Honda\altaffilmark{2,3}, Wako Aoki\altaffilmark{2,3},
Hiroyasu Ando\altaffilmark{2},
Hideyuki Izumiura\altaffilmark{4}, Toshitaka Kajino\altaffilmark{2,3},
Eiji Kambe\altaffilmark{5}, Satoshi Kawanomoto\altaffilmark{2} Kunio
Noguchi\altaffilmark{2,3}, Kiichi Okita\altaffilmark{4}, Kozo
Sadakane\altaffilmark{6}, Bun'ei Sato\altaffilmark{2,7}, Masahide
Takada-Hidai\altaffilmark{8}, Yoichi Takeda\altaffilmark{9}, Etsuji
Watanabe\altaffilmark{4}, Timothy C. Beers\altaffilmark{10} John
E. Norris\altaffilmark{11}, and Sean G. Ryan\altaffilmark{12}}

\altaffiltext{2}{National Astronomical Observatory, Mitaka, Tokyo,
181-8588, Japan; e-mail: honda@optik.mtk.nao.ac.jp, aoki.wako@nao.ac.jp,
ando@optik.mtk.nao.ac.jp, kajino@nao.ac.jp,
kawanomo@optik.mtk.nao.ac.jp, knoguchi@optik.mtk.nao.ac.jp}
\altaffiltext{3}{Department of Astronomy, Graduate University for
Advanced Studies, Mitaka, Tokyo, 181-8588, Japan}
\altaffiltext{4}{Okayama Astrophysical Observatory, National
Astronomical Observatory of Japan, Kamogata-cho, Okayama, 719-0232,
Japan; izumiura@oao.nao.ac.jp, okita@oao.nao.ac.jp, watanabe@oao.nao.ac.jp}
\altaffiltext{5}{Department of Earth and Ocean Sciences, National
Defense Academy, Hashirimizu 1-10-20, Yokosuka, Kanagawa 239-8686,
Japan; kambe@nda.ac.jp}
\altaffiltext{6}{Astronomical Institute, Osaka Kyoiku University,
Kashiwara-shi, Osaka, 582-8582, Japan; sadakane@cc.osaka-kyoiku.ac.jp}
\altaffiltext{7}{Department of Astronomy, School of Science, The
University of Tokyo, Bunkyo-ku, Tokyo 113-0033, Japan; satobn@oao.nao.ac.jp}
\altaffiltext{8}{Liberal Arts Education Center, Tokai University, 1117
Kitakaname, Hiratsuka-shi, Kanagawa, 259-1292, Japan; hidai@apus.rh.u-tokai.ac.jp}
\altaffiltext{9}{Komazawa University, Komazawa, Setagaya, Tokyo,
154-8525, Japan; takedayi@cc.nao.ac.jp}
\altaffiltext{10}{Department of Physics and Astronomy, Michigan State
University, East Lansing, MI 48824--1116; beers@pa.msu.edu}
\altaffiltext{11}{Reseach School of Astronomy and Astrophysics, The
Australian National University, Private Bag, Ewston Creek Post Office,
Canberra, ACT 2611, Australia; jen@mso.anu.edu.au}
\altaffiltext{12}{Department of Physics and Astronomy, The Open
University, Walton Hall, Milton Keynes, MK76AA, UK; s.g.ryan@open.ac.uk}
\altaffiltext{1}{Based on data collected at Subaru Telescope,
which is operated by the National Astronomical Observatory of Japan.}

\begin{abstract}

We have obtained high-resolution ({\it R} $\simeq$ 50,000 or 90,000),
high-quality (S/N $\ga$ 100) spectra of 22 very metal-poor stars ([Fe/H] $\la$
--2.5) with the High Dispersion Spectrograph fabricated for the 8.2m Subaru
Telescope. The spectra cover the wavelength range from 3500 to 5100~{\AA};
equivalent widths are measured for isolated lines of numerous elemental
species, including the $\alpha$ elements, the iron-peak elements, and the light
and heavy neutron-capture elements. Errors in the measurements and comparisons
with previous studies are discussed. These data will be used to perform detailed
abundance analyses in the following papers of this series. Radial velocities
are also reported, and are compared with previous studies. At least one
moderately r-process-enhanced metal-poor star, HD 186478, exhibits evidence of
a small-amplitude radial velocity variation, confirming the binary status noted
previously. During the course of this initial program, we have discovered a
new moderately r-process-enhanced, very metal-poor star, CS~30306--132 ([Fe/H]
$= -2.4$; [Eu/Fe] $= +0.85$), which is discussed in detail in the companion
paper.
 
\end{abstract}


\section{Introduction}

Very metal-poor stars ([Fe/H] $\la -2.5$)\footnote{We use the usual notation
[A/B]$\equiv \log_{10} (N_{\rm A}/N_{\rm B})_* -
\log_{10}(N_{\rm A}/N_{\rm B})_\odot$ and log$\epsilon({\rm A})\equiv
\log_{10}(N_{\rm A}/N_{\rm H})+12.0$, for elements A and B. Also, the term
``metallicity'' will be assumed here to be equivalent to the stellar [Fe/H]
value.} are believed to have been born in the early Galaxy; their chemical
compositions are living records of the nucleosynthesis processes that preceded
their formation. As a result of considerable efforts by many astronomers, a
large list of candidate stars with [Fe/H] $< -2.5$ have been provided by
wide-field objective-prism surveys in the past two decades (e.g., the HK survey:
Beers et al. 1985, 1992; Beers 1999, and the Hamburg/ESO Survey: Christlieb \&
Beers 2000; Christlieb et al. 2001; Christlieb 2003). Over the past several
years, high-resolution spectroscopic studies have enabled the measurement of
elemental abundances for many of the metal-poor stars found by these surveys
(e.g., McWilliam et al. 1995b; Ryan, Norris, $\&$ Beers 1996; Burris et al.
2000; Carretta et al. 2002; Cayrel et al. 2004), including detailed studies of
the lowest metallicity stars yet identified (e.g., Norris, Ryan, \& Beers 2001;
Christlieb et al. 2002). These observational studies, which continue at present,
are providing strong constraints on models of the dominant nucleosynthesis
processes in the earliest epochs of star formation in our Galaxy, in particular
those associated with massive stars and Type II supernovae.

Remarkable progress has been made, in particular, through studies of the
neutron-capture elements in very metal-poor stars. High-resolution spectroscopic
studies of very metal-poor stars have revealed, for example, that a small
fraction (presently estimated to be on the order of 2\%--3\%, Beers, private
communication) of giants with [Fe/H] $< -2.5$ exhibit large overabundances
(e.g., [r-process/Fe] $ > +1.0$) of neutron-capture elements associated with the
r-process (e.g., [r-process/Fe] $ > +1.0$ ;McWilliam et al. 1995b;
Sneden et al. 2000, 2003; Cayrel et al. 2001; Hill et al. 2002).
These, along with a handful of other metal-poor stars with moderately
enhanced r-process elements ($+0.5 \leq$ [r-process/Fe] $\leq +1.0$,
e.g., Westin et al. 2000; Johnson \& Bolte 2001; Cowan et al. 2002),
display remarkably similar abundance patterns in the range 56 $\leq Z <$
76, all apparently in good agreement with the solar-system r-process component.
In addition, some of the neutron-capture-enhanced, metal-poor stars
exhibit abundance patterns associated with s-process nucleosynthesis
(e.g., Norris et al. 1997; Van Eck et al. 2001; Aoki et al. 2002;
Lucatello et al. 2003). 

These efforts are having a large collective impact on studies on
the origin of the neutron-capture elements in the Galaxy
\citep[e.g.,][]{ishimaru99,fields02,qian02}, and on the underlying
physics and astrophysical sites of the r- and s- processes
\citep[e.g.,][]{gallino98,wanajo02,wanajo03,schats02,truran02}.
Furthermore, detailed studies of the r-process-enhanced, very metal-poor stars
have provided new, potentially quite powerful, methods for obtaining hard lower
limits on the age of the Galaxy and the universe, from the application of
cosmo-chronometry based on the observed (present-day) abundance ratios of
radioactive nuclei (Th and U), as compared with one another, and with stable
elements originating in the r-process,
\citep[e.g., Eu,][]{sneden96,wes00,cayrel01,schats02,wanajo02,wanajo03,sneden03}.

In order to develop a more clear understanding of the individual nucleosynthetic
processes that were operating in the early Galaxy, further abundance studies are
required, based on high-quality spectra, for much larger samples of very
metal-poor stars than have been examined to date. We have initiated such a set
of investigations with the Subaru Telescope High Dispersion Spectrograph (HDS,
Noguchi et al. 2002). In this paper we present observations of 22 very
metal-poor stars observed during the commissioning phase of this instrument. In
\S 2 we discuss the selection of targets and details of the
observations that have been carried out. Our spectra cover the wavelength range
from 3500 to 5100~{\AA} with high spectral resolution (a resolving power of $R =
50,000$ or $R = 90,000$) and high signal-to-noise (S/N$ \gtrsim$ 100 per
resolution element). We report the equivalent widths measured for the spectra in
\S 3, where we also discuss the random errors of our measurements, and make
comparisons with previous studies of stars in common. Radial velocity
measurements for our program stars are presented in \S 4, along with a
comparison with previous measurements for a number of stars. These data will be
used in the detailed abundance analyses that will follow in additional papers of
this series.

\section{Observations}
\subsection{Selection of Targets}

The present work is focused primarily on the observed abundance patterns of
r-process elements in very metal-poor stars. Accordingly, our sample was
selected to include stars that fall into one of several categories: (1) Very
metal-poor stars that were previously known to exhibit extremely large
enhancements of their r-process elements (CS~22892--052 and CS~31082--001:
Sneden et al. 1996; Cayrel et al. 2001); (2) Bright metal-poor stars that were
studied by previous authors (e.g., McWilliam et al. 1995b; Burris et al. 2000),
and shown to be moderately r-process-element-rich; (3) Candidate very metal-poor
giants discovered in the course of the HK survey of Beers and colleagues
\citep{beers92,boni00,prieto00}. 
For the majority of these stars, no elemental abundance
results based on high-resolution spectroscopy has been previously obtained. 
Due to the selection criteria employed, it should be noted that our
sample emphasizes stars that are either definite, or suspected,
r-process-enhanced, metal-poor stars, which will impact the
discussion of the distribution of the observed abundances of neutron-capture
elements for these stars presented in Honda et al. (2003; Paper II). 
 
Since our primary purpose is to investigate the neutron-capture elements, we
selected giants, whose metal lines are generally stronger than metal-poor dwarfs
near the main-sequence turnoff due to their lower effective temperatures.
Exceptions are HD~140283 and BS~17583--100, which were observed for comparison
purposes. A rather large fraction of very metal-poor stars exhibit enhancements
of carbon (e.g., Beers et al. 1992, Rossi et al. 1999), up to 25\% by some
recent estimates. However, strongly carbon-enhanced stars ([C/Fe] $\ga$ +1.0)
are excluded from our sample, because contamination arising from molecular lines
(CH and CN) makes the analysis of lines of neutron-capture elements difficult,
and causes particular problems with regard to features of Th and U. An exception
is the star CS~22892--052, which is known to exhibit an extremely large excess of
r-process elements, and a large carbon enhancement, on the order of [C/Fe]
$\approx +1.1$ (see Norris, Ryan, \& Beers 1997 for a discussion of the impact on
studies of Th in such stars).

The 22 stars selected for our program are listed in Table~\ref{tab:obs}. In this
table we also list the apparent $V$ magnitudes and $B-V$ colors, taken from the
list of Beers et al. (2003 in preparation) and the SIMBAD database. As can be seen,
most of our targets fall in the range $0.7 \le B-V \le 1.2$, and are likely to
be giant-branch stars.

\subsection{Subaru/HDS Observations}

High-resolution spectra of our program stars were obtained during the
commissioning phase of HDS between July 2000 and July 2001 -- a detailed log is
provided in Table~\ref{tab:obs} and Table~\ref{tab:obs2}. The HDS detector is a mosaic
system of two EEV-CCDs, each with 2048 $\times$ 4100 pixels. HDS is designed to
achieve high spectral resolving power, high sensitivity, and (almost) complete
wavelength in the blue region. These are essential characteristics for our
program, since the weak absorption lines of neutron-capture elements fall
primarily in the near UV--blue range. Details of the design of the spectrograph
and its performance are provided by \citet{nogu02}.

For the observations reported herein, the slit width of the spectrograph was set
to 0.4 arcsec (200 $\mu$m) or 0.72 arcsec (360 $\mu$m), which corresponds to a
spectral resolving power of {\it R} $\simeq$ 90,000 or 50,000, respectively
(Table~\ref{tab:obs}). An exception is HD~186478, which was observed with a 0.36
arcsec (180 $\mu$m) slit. The high resolving power and oversampling of the
spectra (roughly six pixels per resolution element for 0.72 arcsec slit width)
obtained by HDS are particularly valuable for the study of lines affected by
hyperfine splitting and isotope shifts, and/or from blending with other atomic
and molecular lines. Our observations covered the wavelength from 3500 {\AA}
(3400 {\AA} for a few objects) to 5200 {\AA} (5100 {\AA} for a few stars), with
a lack of data between 4350 {\AA} and 4400 {\AA} (4230 {\AA} and 4280 {\AA}) due
to the gap between the two CCDs.

Since our observations were made in the early phase of HDS commissioning, there
were some limitations of various components in the pre-slit unit, especially the
image rotators and the atmospheric differential dispersion corrector (ADC). The
ADC was installed at the end of 2000, and has been applied from the January 2001
run forward. An image rotator was needed for target acquisition and guiding in
the observing run conducted in 2000. In the July 2000 run, the blue spectra of
HD~115444, HD~122563, and HD~140283 were obtained using the image rotator
optimized for the red, as the blue one was unavailable at that time. In
addition, the slit was fixed to the north/south direction, instead of being
aligned along the parallactic angle, due to limitations in guiding. For this
reason, significant light loss occurred in the short-wavelength region where the
effect of atmospheric differential dispersion is quite large, degrading the
spectral quality at the shortest wavelengths. We note that these three objects
were re-observed in later observing runs so that sufficient quality could be
achieved. 

In most cases, the spectra of fainter stars in our program were obtained by
combining several 1800 sec exposures. This choice was motivated by the desire to
limit the degradation of the spectra due to cosmic ray events. The total
exposure time for each object ranges from 900 sec (for the brightest star) to
9751 sec (for the faintest star). The signal-to-noise ratios at $\sim$ 4000
{\AA} are 40 $ < $S/N$ < $ 450 per pixel (100 $ < $S/N$ < $ 900 per resolution
element), as shown in Table~\ref{tab:obs}. There was no need, in general, for
the use of on-chip CCD binning, because the read-out noise was not an important
source of noise in this study. The exception is CS~31082--001, for which
2$\times$2 binning mode was used, because this object was observed during
another observing program in which the binning mode was employed.

For reduction of the spectral data, we obtained bias frames, halogen lamp frames
for flat-fielding, and Th-Ar spectra for wavelength calibration. Though dark
frames were obtained in each run to check the dark current of the CCD, it turned
out to be very small, hence no dark correction was made during data reduction.

The echelle data were processed using the IRAF\footnote{IRAF is distributed by
National Optical Astronomy Observatories, which are operated by the Association
of Universities for Research in Astronomy, Inc., under cooperative agreement
with the National Science Foundation.} software package in a standard manner.
Here we summarize the flow of the data reduction. We first corrected the
fluctuation of the bias level by subtracting the average of the counts in the
over-scan region from each frame. The median of the bias frame was then
subtracted from all frames. We then divided the object frames by the average of
the flatfield frames. The scattered light level was estimated by obtaining
surface fits of the inter-order regions, and then was subtracted from each
object frame. One-dimensional spectra were extracted after removing cosmic ray
events. Since the wavelength ranges covered by the CCD are much wider than the
free spectral range in the near UV-Blue region, we trimmed the spectral orders
when the count level fell below a useful level. In practice, this meant that
roughly 1000 pixels were trimmed from the blue portions of each order, and 500
pixels were trimmed from the red portions of each order, respectively.

The S/N ratios of the spectra were evaluated from the peak count in echelle
order 149 ($\sim$4000 {\AA}). The S/N ratios per pixel (0.012 {\AA}) and per
resolution element are given in Table~\ref{tab:obs}. It should be noted that the
sampling rate of HDS is quite high in the $R\simeq50,000$ spectra (six pixels per
resolution element), hence the S/N ratios per pixel may seem rather low in
some cases.

\begin{figure}
\begin{center}
\includegraphics[width=13cm]{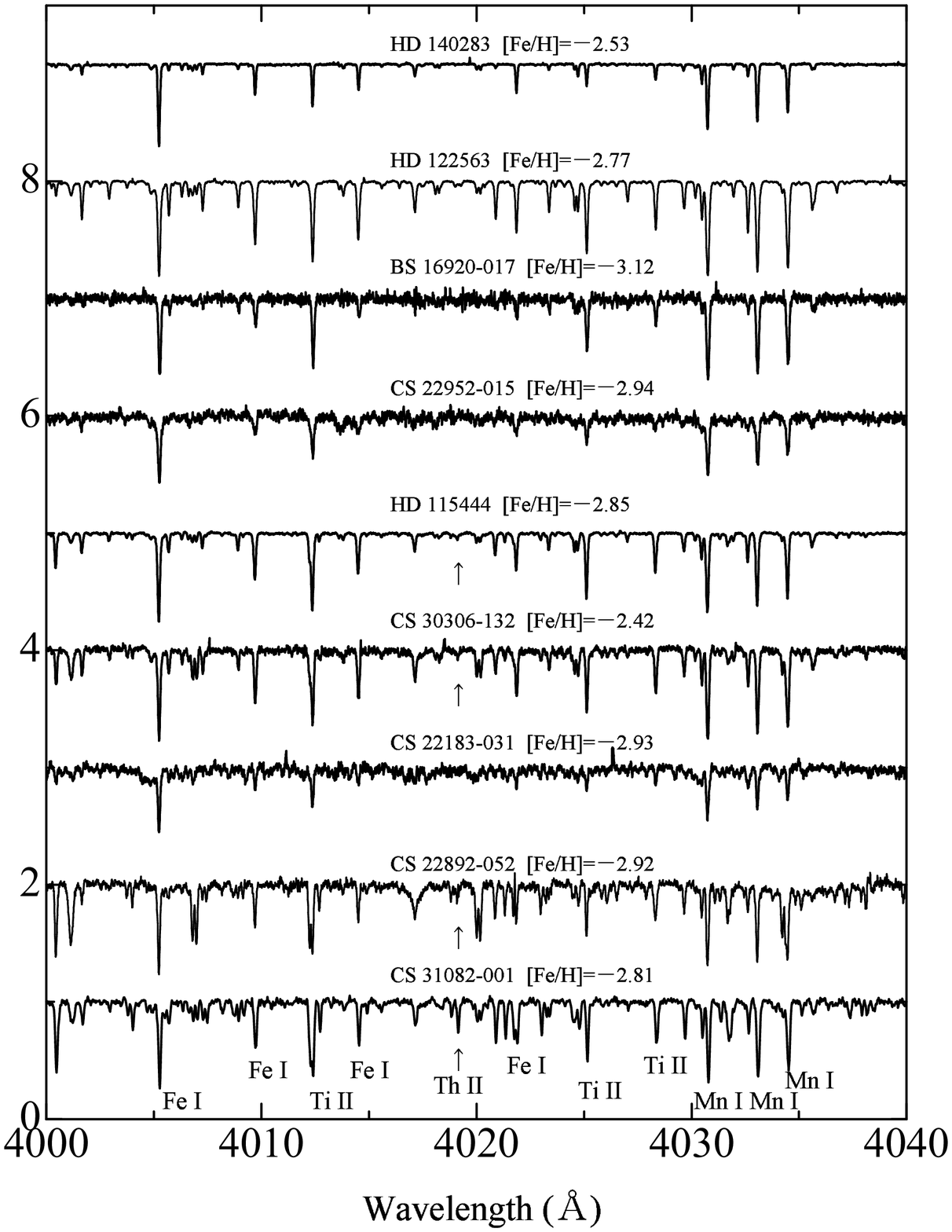}
\caption{Examples of spectra for nine stars in our sample over the
wavelength range 4000--4040{\AA}. The star names and [Fe/H] values reported
in Paper II are presented in the figure. This spectral region 
includes the line of \ion{Th}{2} at 4019~{\AA}; detections are indicated
with a vertical arrow. The apparent emission lines seen in some of the
 spectra result from imperfect cosmic-ray cleaning.}
\label{fig:spec1}
\end{center}
\end{figure}

\begin{figure}
\begin{center}
\includegraphics[width=13cm]{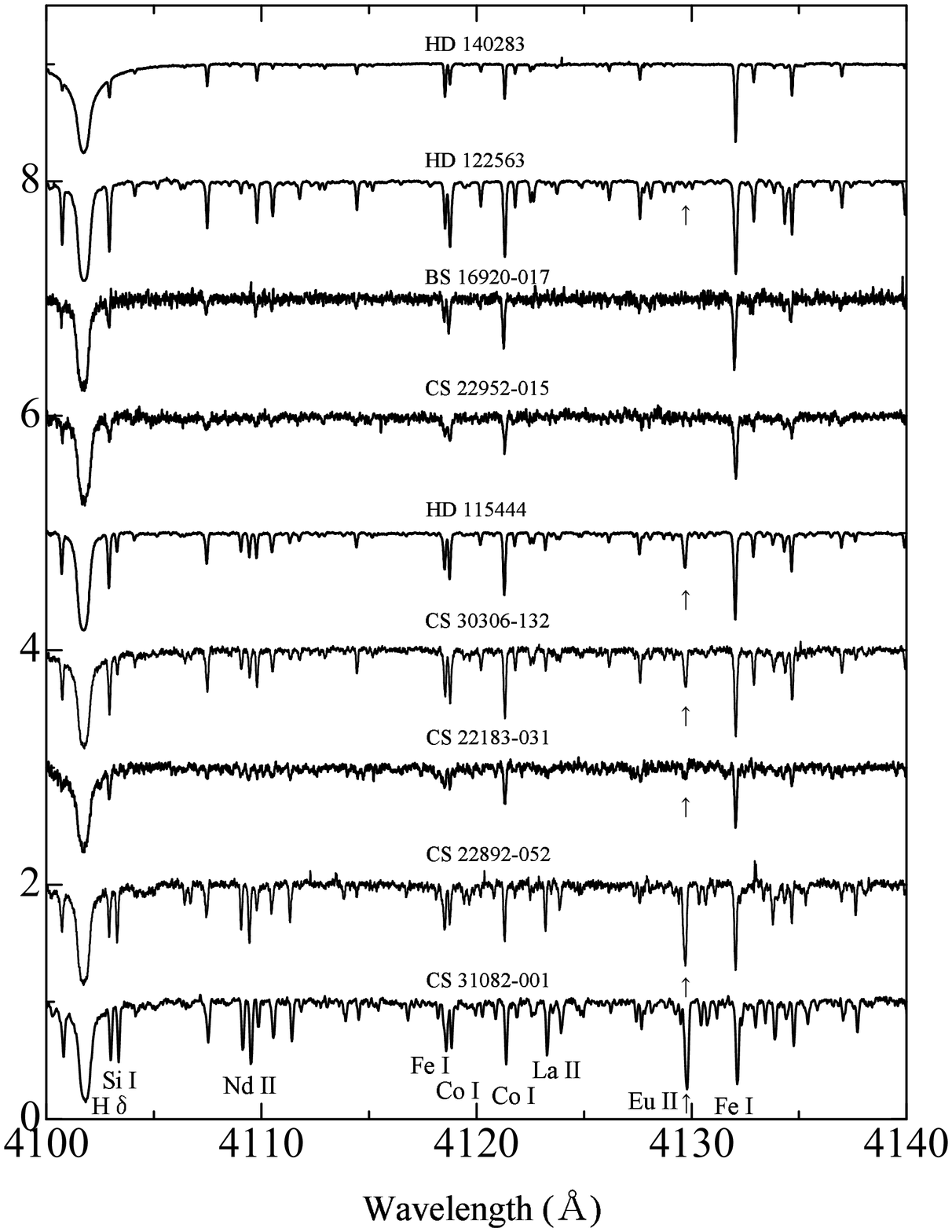}
\caption{The same as Figure \ref{fig:spec1}, but for the
spectral range 4100--4140{\AA}. This region includes lines of H$\delta$ (from
which one ascertain that the effective temperatures of these stars are quite
similar to one another) and 
the  \ion{Eu}{2} 4129~{\AA} line; detections are indicated with a vertical
arrow. 
}
\label{fig:spec2}
\end{center}
\end{figure}

Sample spectra for nine of our program stars in the regions near 4000 {\AA}
(which includes the \ion{Th}{2} 4019 \AA\ line) and 4100 {\AA} (which includes
the \ion{Eu}{2} 4129 \AA\ line) are shown in Figures~\ref{fig:spec1} and
~\ref{fig:spec2}, respectively. HD~140283 and HD~122563 are familiar,
well-studied metal-poor stars. BS~16920--017 is the star with the lowest
metallicity in our study, with [Fe/H] $= -3.1$. CS~22952--015 has been studied
by previous authors \citep{mcwi95b,ryan96}. In the spectrum of CS~22183--031,
the \ion{Eu}{2} 4129 {\AA} feature is detected, but it is quite weak. Given the
paucity of information on the Eu abundances of stars with [Fe/H] $\sim -3.0$,
this object should be re-observed at higher S/N in order to obtain a better
measurement. The other four objects in these figures show enhancements of the
r-process elements. CS~22892--052 and CS~31082--001 are well-known, extremely
r-process-enhanced, very metal-poor giants \citep{sneden96,cayrel01}. These two
objects clearly show the \ion{Th}{2} 4019 {\AA} line, as well as very strong
\ion{Eu}{2} 4129 {\AA} features. HD~115444 is the object shown by \citet{wes00}
to exhibit a moderate excess of r-process elements. CS~30306--132 turned out to
exhibit excesses of the neutron-capture elements, as discovered during the
present work. The \ion{Th}{2} 4019 {\AA} line was clearly detected in this object (see
Paper II for details).

\section{Equivalent Width Measurements}

We identified Fe absorption lines between 3700 {\AA} and 5100 {\AA}, which will
be used to determine the atmospheric parameters for the abundance analysis using
model atmospheres. Identification of these lines was mostly made on the basis of
the line list provided by \citet{wes00}. Equivalent widths were measured for
clear, unblended lines of \ion{Fe}{1} and \ion{Fe}{2} by fitting gaussian
profiles to the observations using the spectral analysis software SPTOOL,
developed by Y. Takeda (private communication). We excluded lines from our
analysis that may be significantly blended with other absorption lines. The
blending with other atomic lines was checked by using the atomic line list by
Kurucz \& Bell (1995).

Gaussian fitting may not well reproduce the wings of strong lines. However, for
weak lines, the difference in derived equivalent widths from evaluations based
on gaussian fitting and those based on direct integration over the observed line
profile is very small. Since our analysis relies exclusively on weak lines, we
measured all equivalent widths by the gaussian fitting procedure.

In addition to Fe lines, we also identified absorption lines of other elements,
using the line lists provided by \citet{wes00} and \citet{sneden96}.
For Ba lines, we adopt the list of McWilliam et al. (1998).
For La, Eu, and Tb lines, we take from the list of Lawler et al. (2001a,b,c).
We use the newest data also for Nd and Yb which are derived by Den
Hartog et al. (2003) and C. Sneden (private communication).
Equivalent widths of these lines were also measured in the same manner as
applied to the Fe lines. The measured equivalent widths are given in
Table~\ref{tab:ew1}.
The line data (lower excitation potentials, L.E.P., and the $gf$-values)
are also listed in Table~\ref{tab:ew1}.

\subsection{Estimates of Internal Errors}\label{sec:error}

We estimate the random (internal) errors of our derived line strengths by
determining the differences in measured equivalent widths from two
measurements of each spectrum obtained with different individual exposures.

\begin{figure}[p]
\begin{center}
\includegraphics[width=6cm]{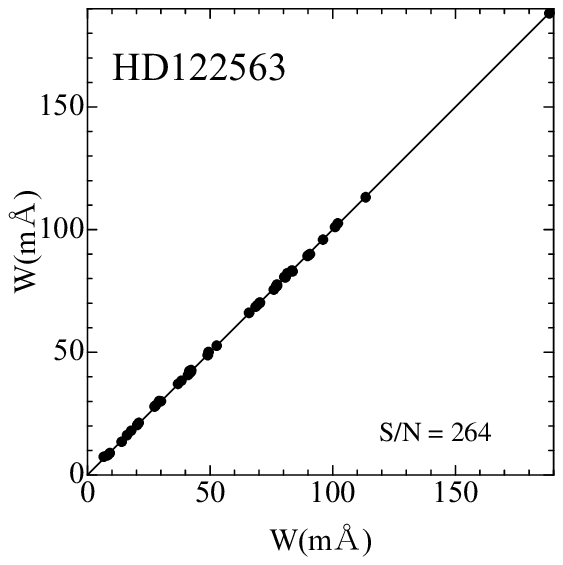}

\includegraphics[width=6cm]{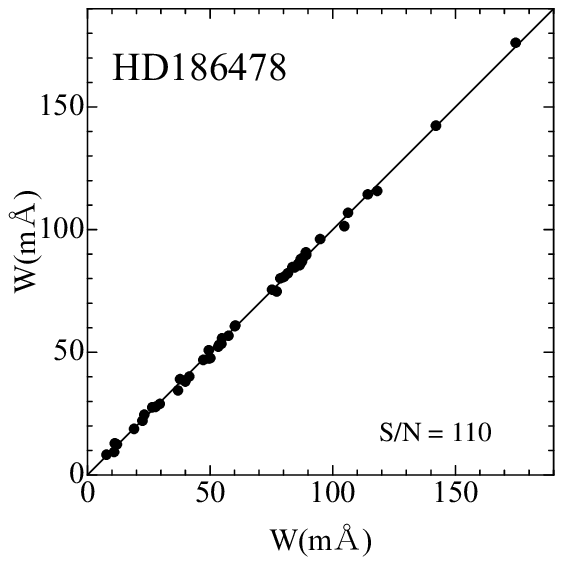}

\includegraphics[width=6cm]{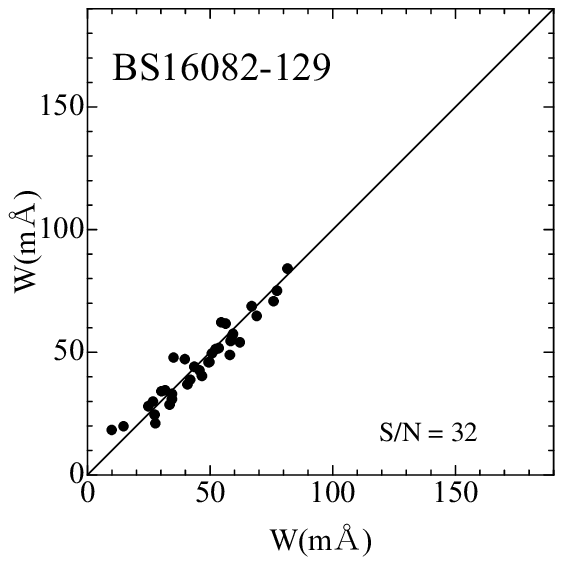} 
\caption{Comparison of equivalent width (W) measurements for
spectra obtained by individual HDS exposures for HD~122563 (top),
HD~186478 (middle), and BS~16082--129 (bottom).  The S/N ratio of the
final summed spectra are given in the individual panels. 
See text for details.}\label{fig:ew} 
\end{center} 
\end{figure}

We selected three objects, HD~122563, HD~186478, and BS~16082--129, as
representive of stars observed with high, moderate, and rather low S/N ratios,
respectively. The number of photons collected by each exposure is about 70000,
12000, and 1000 at 4000~{\AA} for HD~122563, HD~186478, and BS~16082--129,
respectively. Comparisons between the two measurements of weak Fe lines for
these three objects are shown in Figure~\ref{fig:ew}. No systematic differences
between the individual measurements are evident. The standard deviations in the
differences of the two measurements for HD~122563, HD~186478, and BS~16082--129
are 0.39, 1.26, and 4.98 m{\AA}, respectively.

The uncertainly in the measured equivalent widths may also be roughly estimated,
based on the S/N ratio of the spectrum, as (line width) $\times$ (S/N)$^{-1}$.
The typical line widths for giant stars in our sample is 7.5~km~s$^{-1}$
(100~m{\AA} at 4000~{\AA}). The uncertainty expected from the S/N ratio of each
spectrum, which is taken to be the square-root of the number of detected
photons, is 0.38, 0.90, and 3.2 m{\AA} for HD~122563, HD~186478, and
BS~16082--129, respectively. The random errors measured above show a reasonable
agreement with these values, though the measured ones are slightly higher than
those predicted from the S/N ratios. This small discrepancy may arise because
the Fe lines used in our analysis are randomly distributed across individual
orders, which have a blaze function variation in their S/N levels, while the
number of photons was measured at the center of the echelle blaze profile.

\subsection{Comparisons with Previous Studies}\label{sec:comp}

\begin{figure}[p]
\begin{center}
\includegraphics[width=5cm]{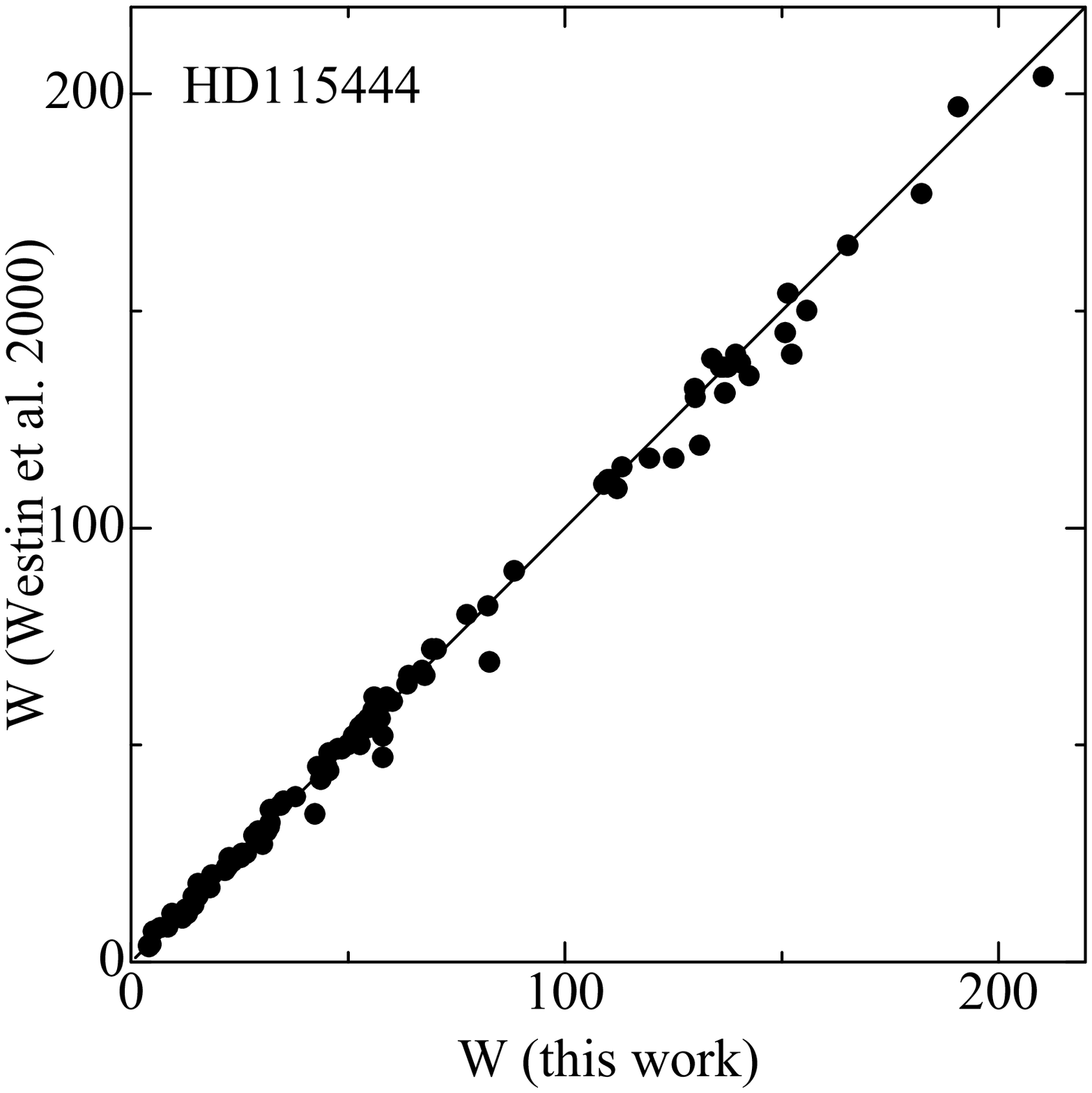}
~~~~~
\includegraphics[width=5cm]{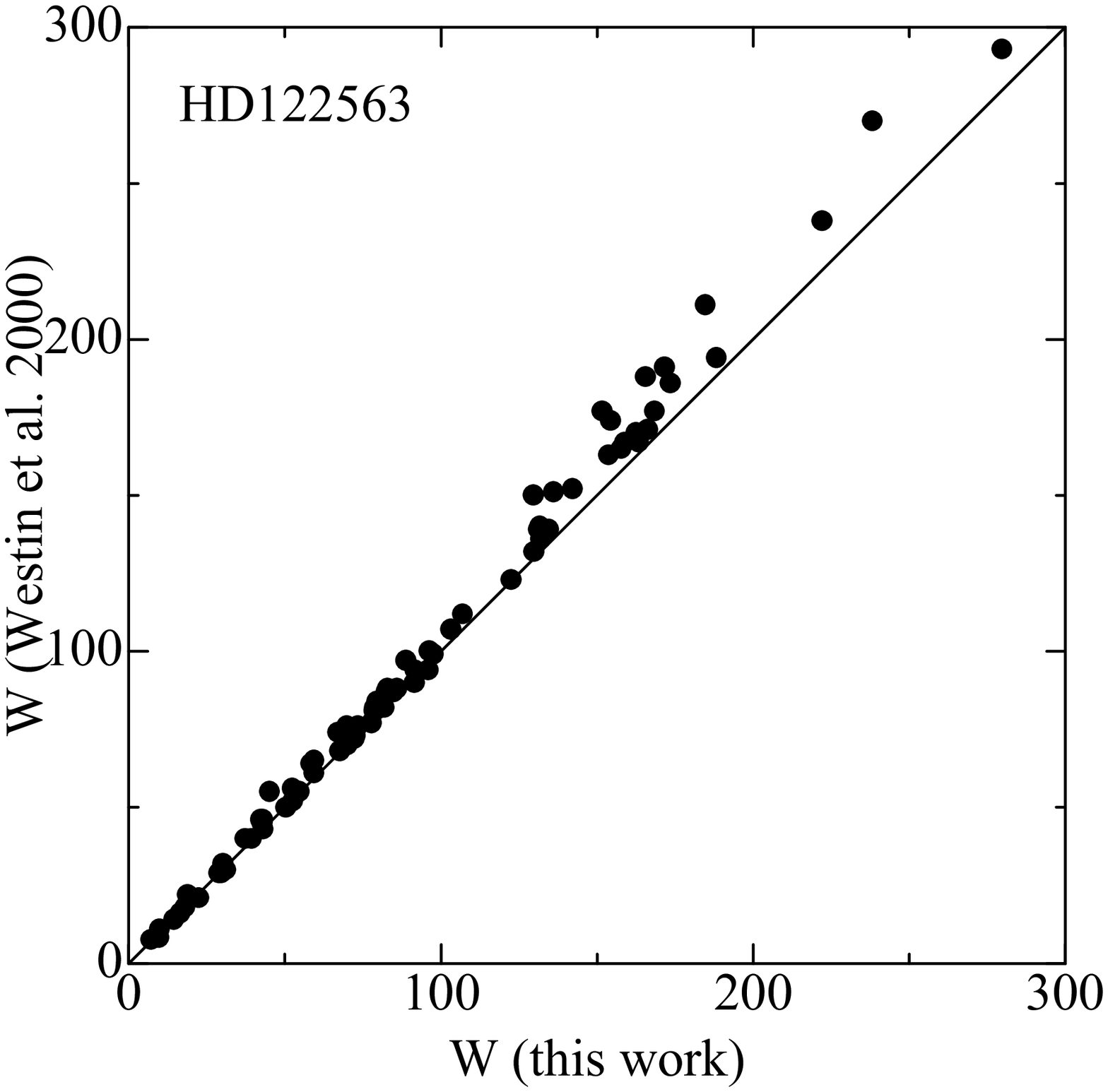}

\caption{Comparison of equivalent width (W) measurements (m{\AA}) by
Westin et al. (2000) and this work.}\label{fig:westin}
\end{center}
\end{figure}

\begin{figure}[p]
\begin{center}
\includegraphics[width=5cm]{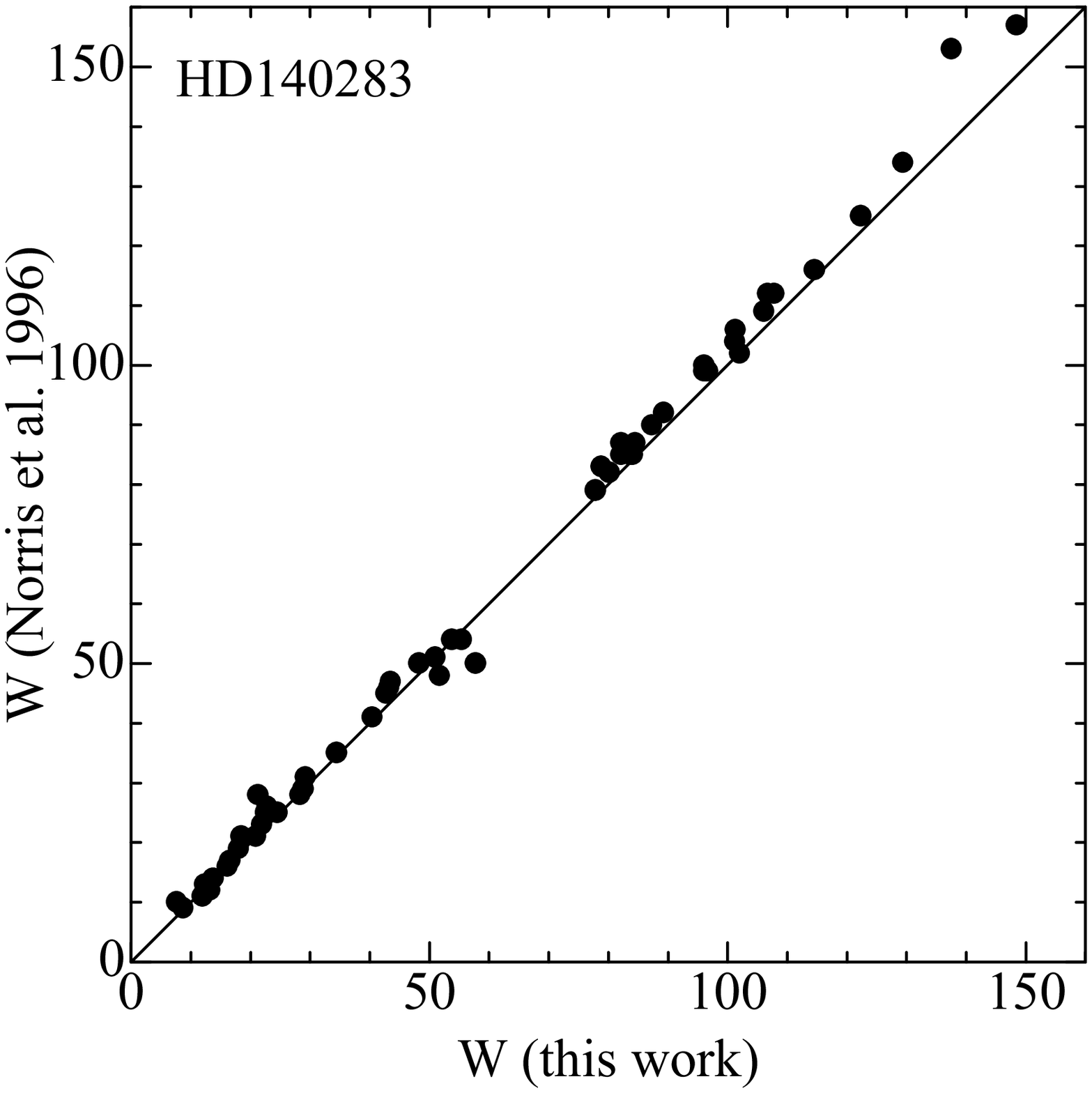}
~~~~~
\includegraphics[width=5cm]{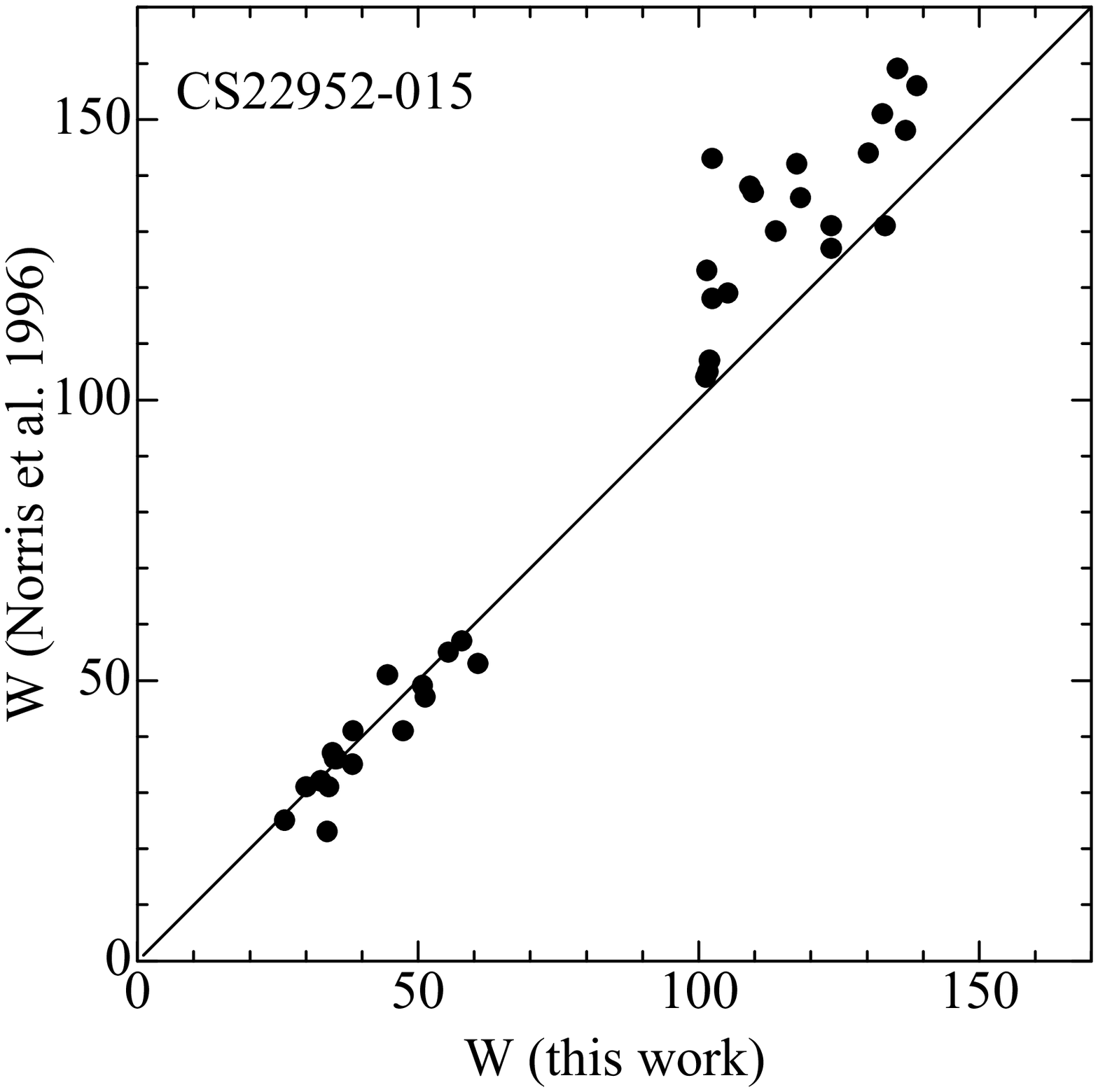}
\caption{Comparison of equivalent width (W) measurements (m{\AA}) by Norris et al. (1996) and this work.}\label{fig:norris}
\end{center}
\end{figure}

\begin{figure}[p]
\includegraphics[width=4cm]{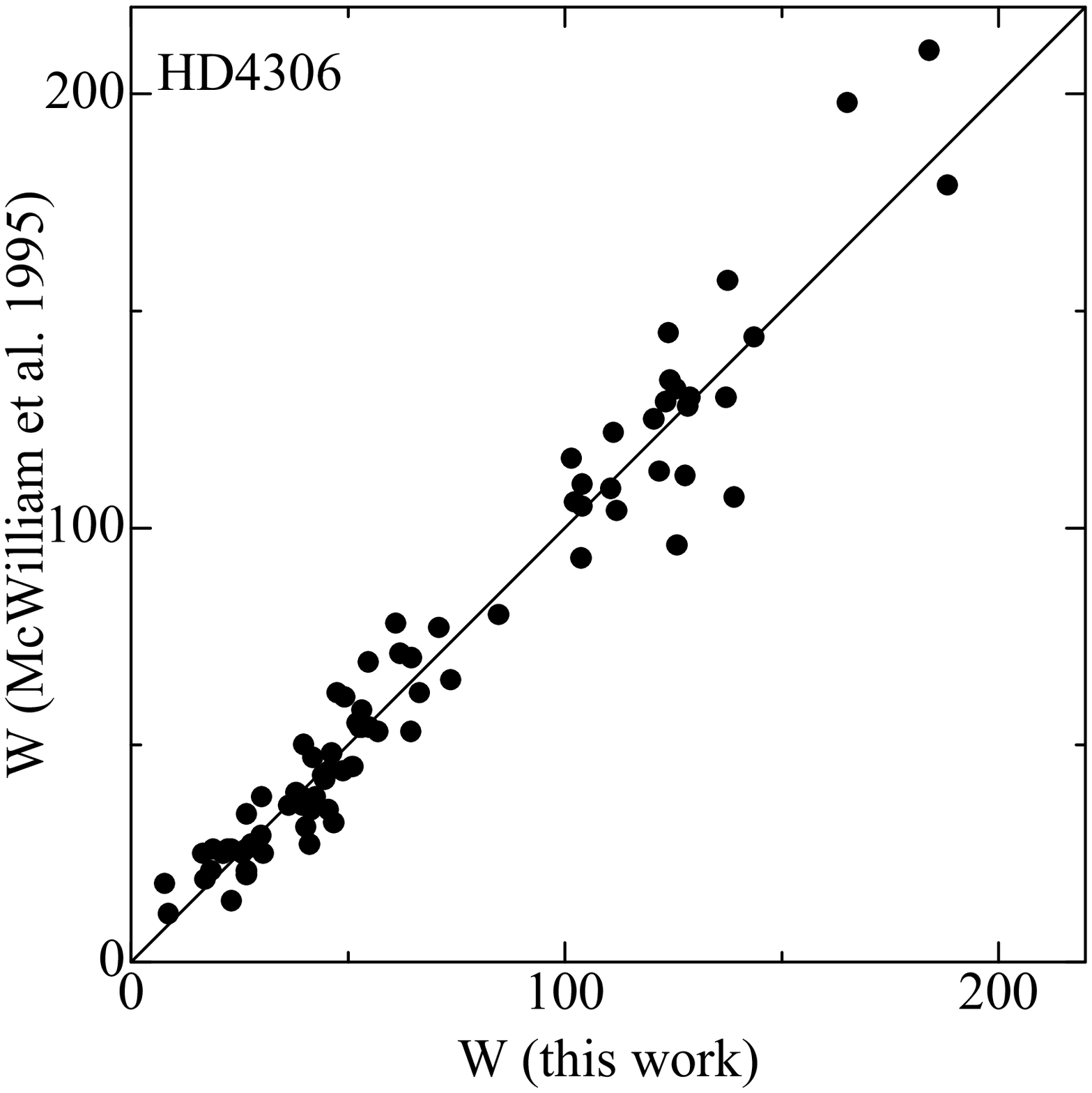}
~~~~
\includegraphics[width=4cm]{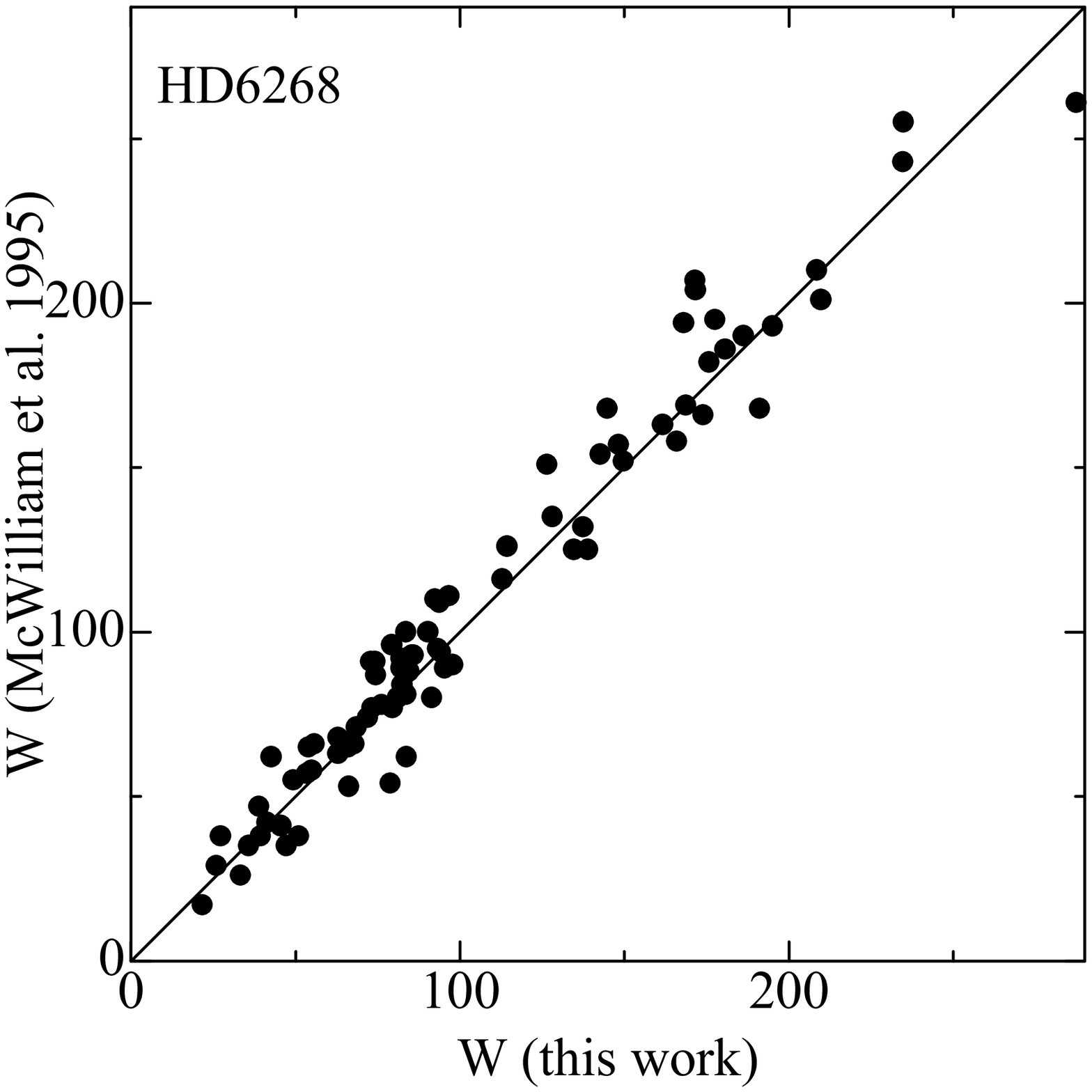}
~~~~~~~~

~

~

\includegraphics[width=4cm]{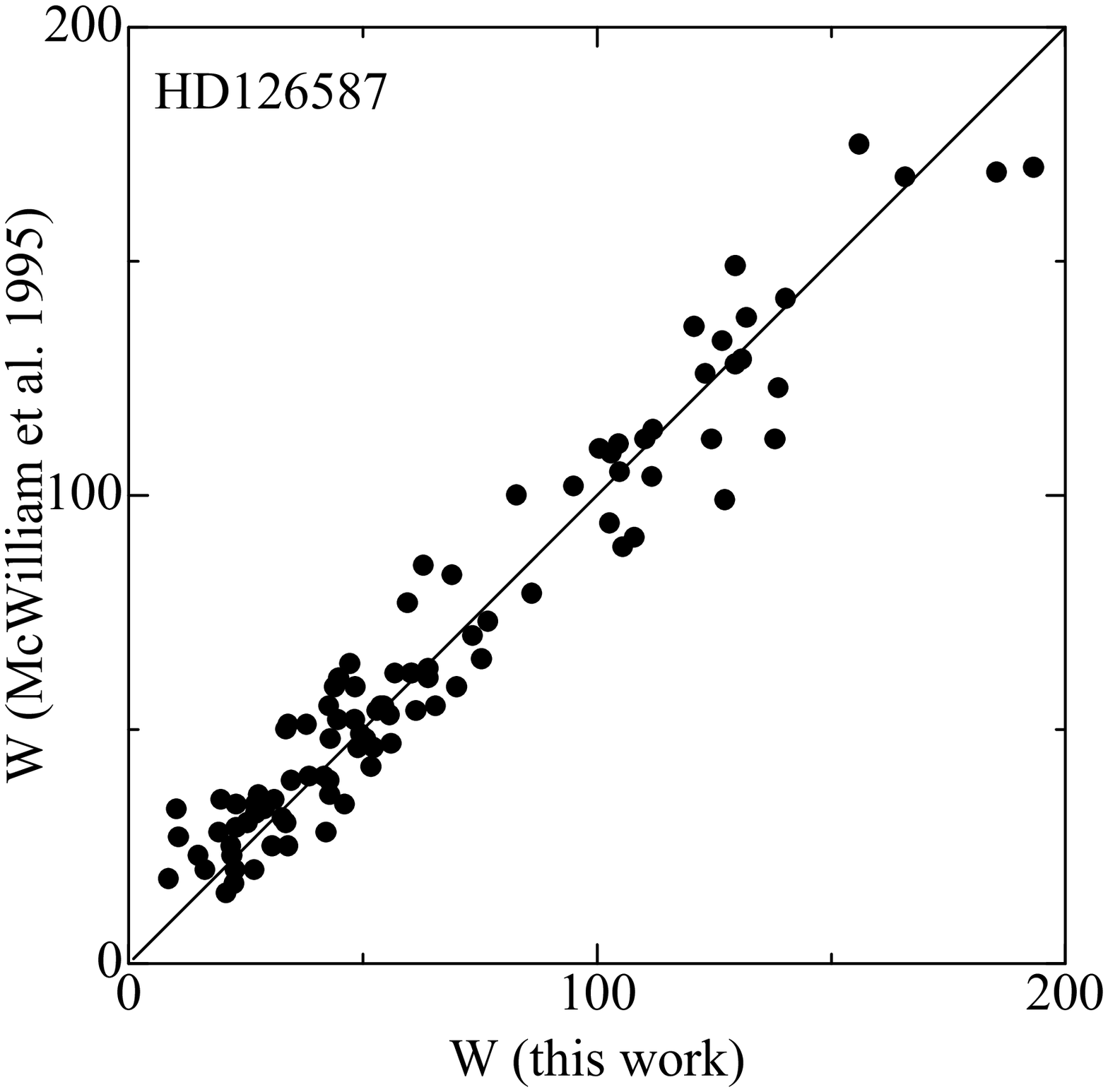}
~~~~
\includegraphics[width=4cm]{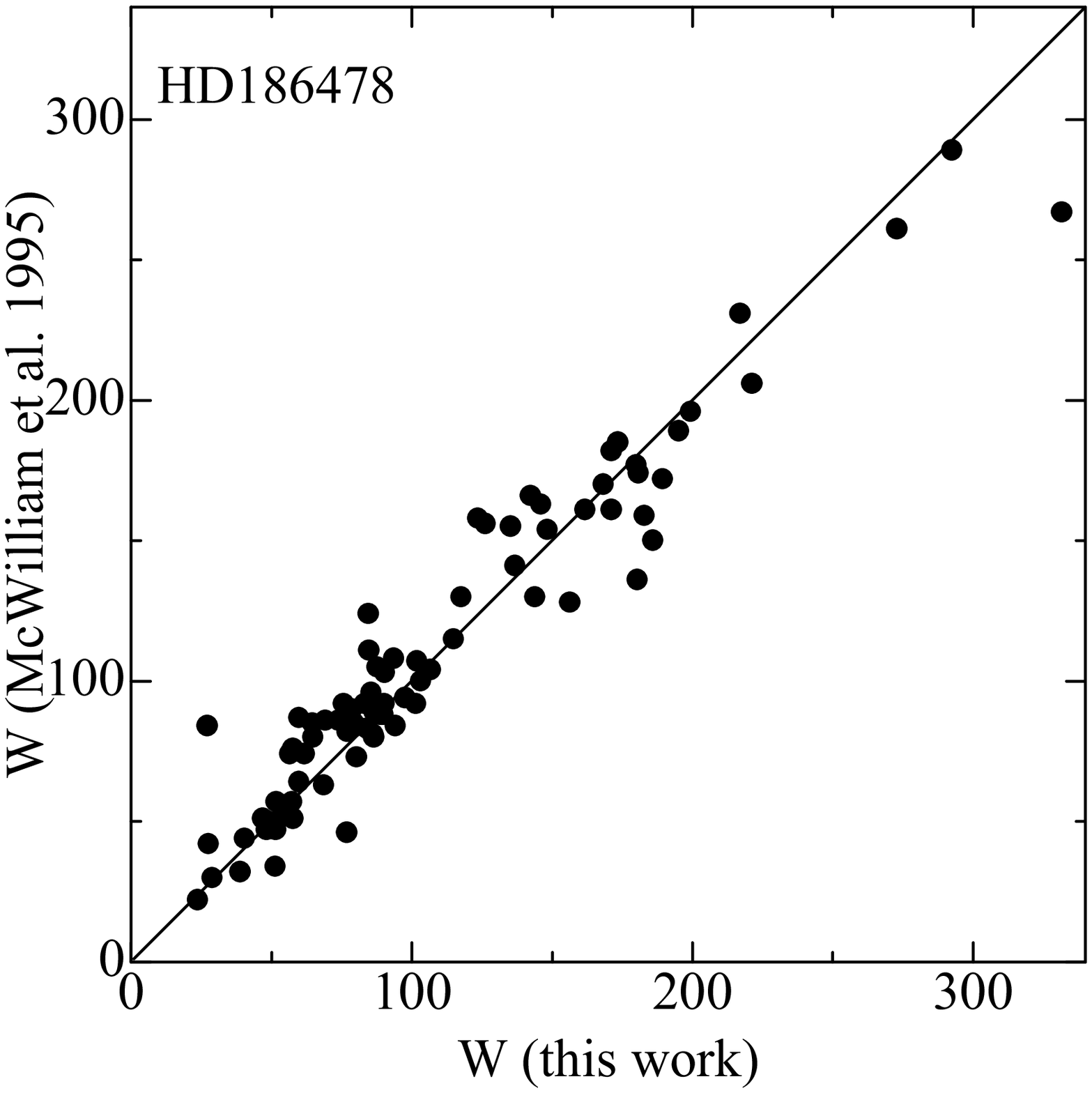}
~~~~~~~~

~

~

\includegraphics[width=4cm]{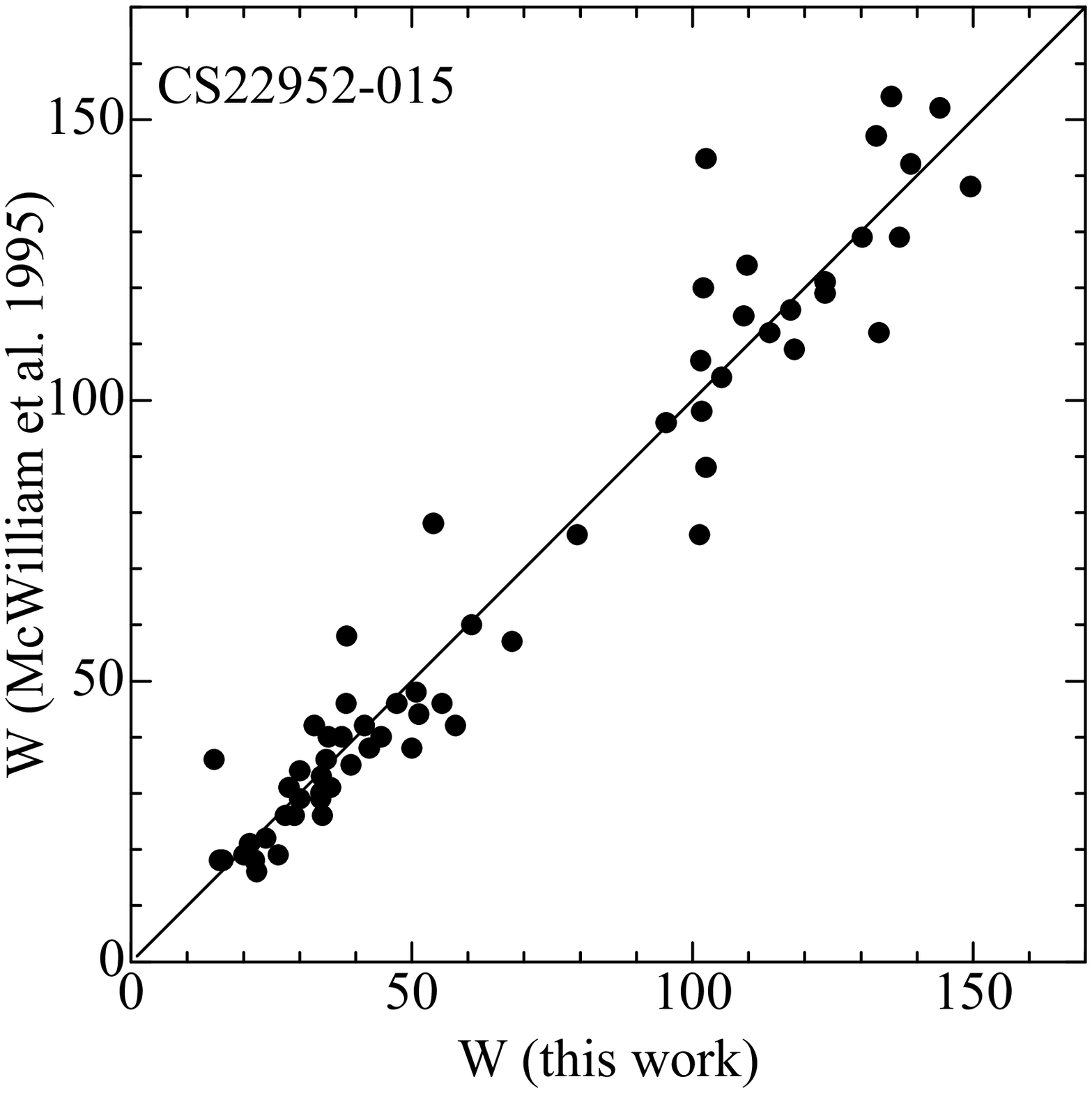}
~~~~
\includegraphics[width=4cm]{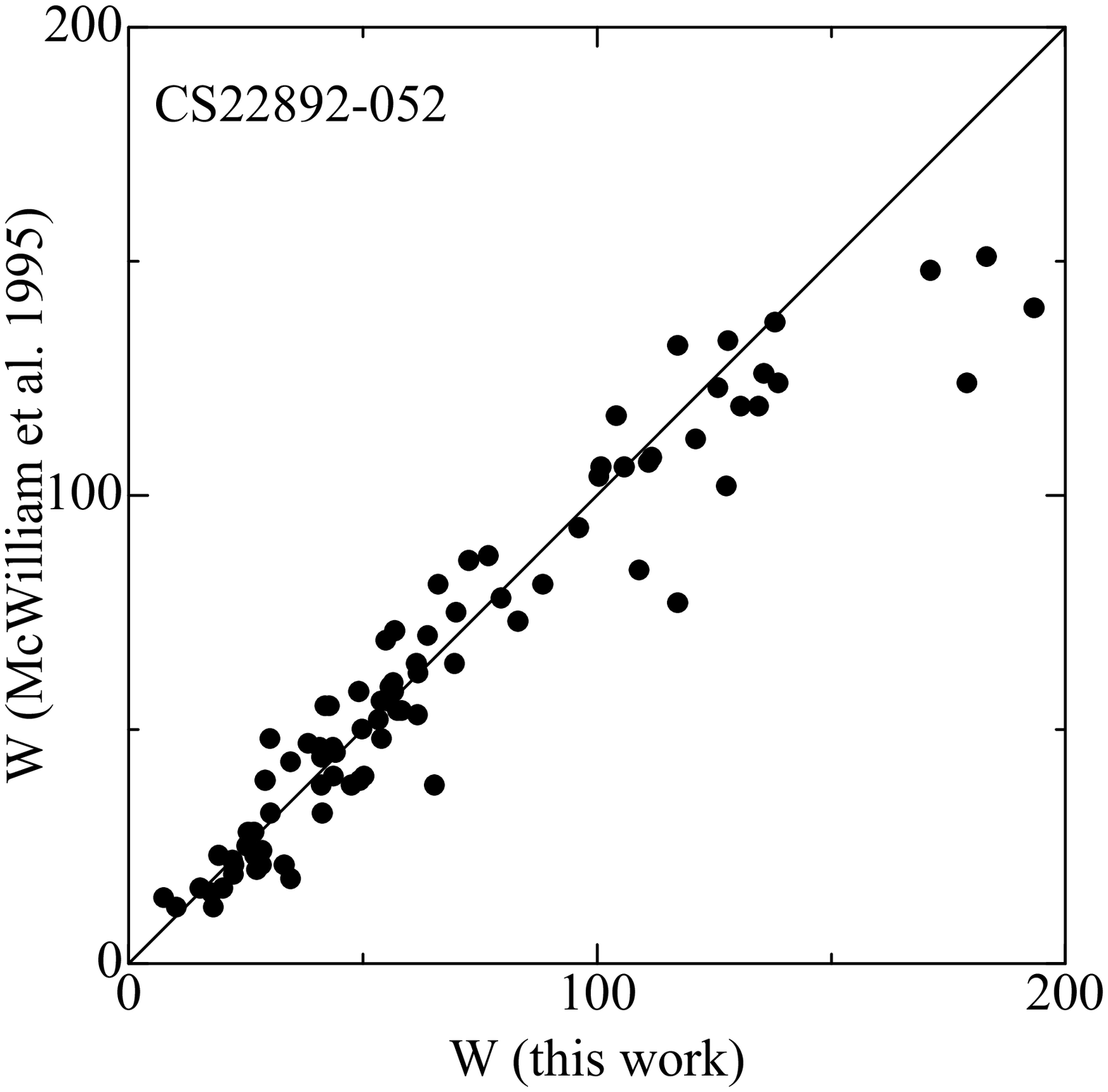}
~~~~~~~~

~

~

\caption{Comparison of equivalent width (W) measurements (m{\AA}) by McWilliam
 et al. (1995a) and this work.}\label{fig:mcwilliam}
\end{figure}

\begin{figure}[p]
\begin{center}
\includegraphics[width=10cm]{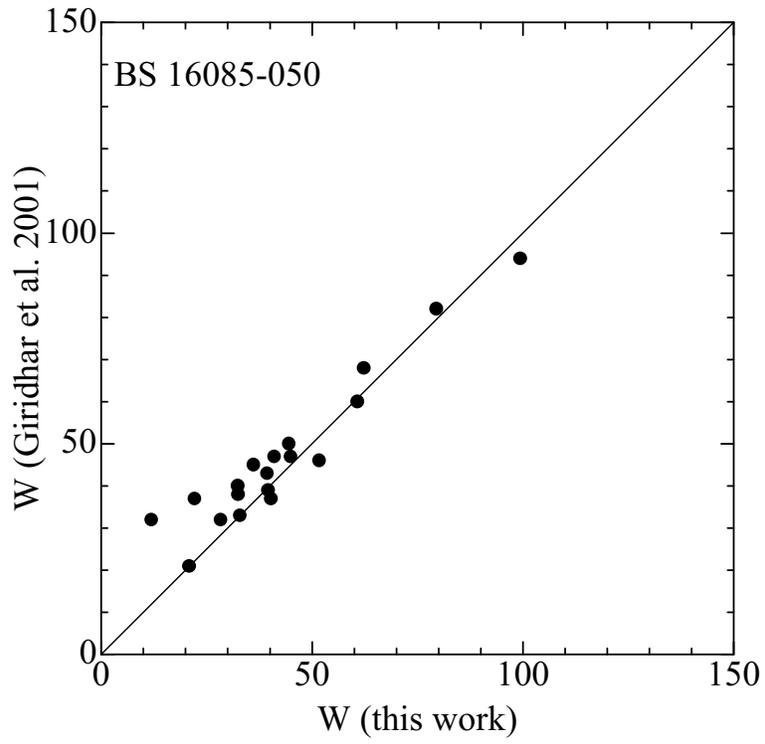}
\caption{Comparison of equivalent width (W) measurements (m{\AA}) by
Giridhar et al. (2001) and this work.}\label{fig:giridhar}
\end{center}
\end{figure}

\begin{figure}[p]
\begin{center}
\includegraphics[width=10cm]{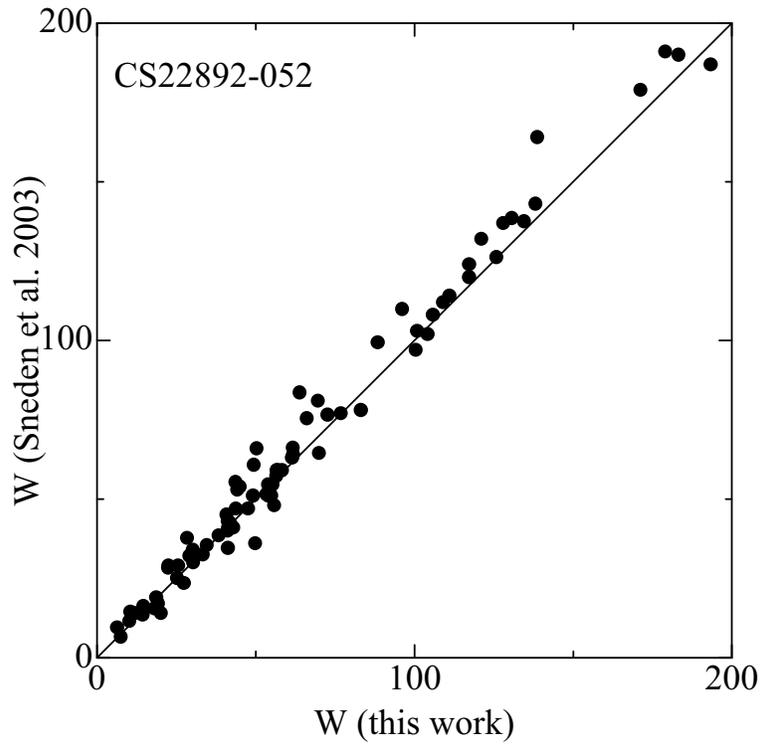}
\caption{Comparison of equivalent width (W) measurements (m{\AA}) by
Sneden et al. (2003) and this work.}\label{fig:sneden}
\end{center}
\end{figure}

Several stars in our sample have also been investigated by previous authors
conducting high-resolution abundance studies. In Figures~\ref{fig:westin}-\ref{fig:sneden}, the equivalent widths estimated from the present data
are compared with those reported by others.

Westin et al. (2000) analyzed high-resolution ($R \sim 60,000$) and high S/N
($\sim $ 200 at 4000 {\AA}) spectra of the two bright objects HD~122563 and
HD~115444, obtained with the ``2d-coud'' cross-dispersed echelle spectrograph at
the McDonald Observatory 2.7m telescope; their results are compared with ours in
Figure~\ref{fig:westin}. An excellent agreement is found, with a very small
scatter ($\sim$ 2$\%$) in the range of equivalent width less than 150 m{\AA}.
There is a small ($\sim 10$\%) difference in the range of equivalent widths
larger than 150 m{\AA} for HD~122563. The reason for this difference is not
clear, but this difference does not have a significant influence on the abundance
analysis because we are primarily concerned with weak lines.

Norris et al. (1996) obtained high-resolution spectra for two of the very
metal-poor stars included in our sample, using the coude spectrograph (UCLES) at
the Anglo-Australian Telescope. They studied spectra with $R\sim$ 40,000 of
CS~22952--015 (S/N $\sim$ 50) and HD~140283 (S/N $\sim$ 200). In
Figure~\ref{fig:norris}, the equivalent widths measured for our spectra are
compared with theirs for these two objects. The agreement is very good for
HD~140283; there is no systematic difference, and the dispersion is quite small
($\sim$ 2$\%$). On the other hand, the scatter in the comparison for
CS~22952--015 is larger, and our equivalent widths are systematically smaller
than those of Norris et al. (1996) ($\sim$ 10$\%$) for lines with large
equivalent widths ($>100$ m~{\AA}). The large scatter is presumably due to the
lower S/N ratios in the CS~22952--015 spectra than those of HD~140283 in both
studies. Most of the lines with equivalent widths stronger than 100 m{\AA} come
from the portions of the spectra at wavelengths blueward of 4000 {\AA}. We
suspect that the discrepancy found for these strong lines is due to errors in
the measurements by \citet{norr96}, which were based on spectra of rather low
S/N in this wavelength region.

In Figure~\ref{fig:mcwilliam}, equivalent widths for HD~4306, HD~6268,
HD~126587, HD~186478, CS~22952--015, and CS~22892--052, reported by McWilliam et
al. (1995a), are compared with ours. Their spectra were obtained with the
2D-Frutti photon-counting imager at the Las Campanas 2.5m telescope. The typical
S/N of the observations obtained by \citet{mcwi95a} is S/N $\sim 40$ with {\it
R} $\sim$ 22,000 at 4800 {\AA}. Their wavelength coverage extends from 3600
{\AA} to 7600 {\AA}. In spite of the large dispersion in the equivalent widths
between these two set of measurements, which is surely due to the low S/N ratios
in the spectra of \citet{mcwi95a}, the equivalent widths of \citet{mcwi95a}
exhibit no systematic difference with respect to ours. One exception is the
comparison with four strong lines in CS~22892--052. We suspect that this
deviation is due to errors in the data of \citet{mcwi95a}, because these lines
exist in the shortest wavelength regions, where the quality of the McWilliam et
al. data is quite low.

In Figure~\ref{fig:giridhar}, equivalent widths for BS~16085--050, reported by
Giridhar et al. (2001), are compared with ours. Their spectra were obtained
with the Apache Point Observatory's 3.5m telescope and vacuum-sealed echelle
spectrograph. Though the comparison shows a rather large dispersion, which is
likely due to the low S/N of the spectrum reported by Giridhar et al. (2001),
there is no systematic difference between the two measurements. Although they
also reported the equivalent widths of CS~22169--035, there are only four iron
lines which were observed by both studies. Therefore we do not show the
comparison for this object.

Recently, Sneden et al. (2003) reported the results of a new detailed analysis
of CS~22892--052. Their optical spectrum was obtained with Keck I/HIRES,
McDonald 2.7m/2d-coude, and VLT/UVES. In Figure~\ref{fig:sneden}, we compare our
measured equivalent widths for CS~22892--052 with theirs. There exists no
systematic difference between the two measurements, and the agreement is better
than that found in the comparison with McWilliam et al. (1995a) for the same
star.

\section{Measurements of Radial Velocities}\label{}

Information on radial velocities and (where proper motions are available) on the
space motion of stars of the halo is required to understand the structure and
formation of the Galaxy. Precise radial velocity measurements are of particular
importance for the moderately and highly r-process-enhanced, metal-poor stars,
in order to check on their possible binarity, as this may impact the likely
astrophysical site(s) of the r-process. For example, Qian \& Wasserburg (2001)
have suggested that the r-process-enhanced, metal-poor stars were produced by
contamination from companions that underwent Type II SN explosions.

Measurements of radial velocities were made for selected clean iron lines used
in the equivalent width measurements. The wavelengths of the lines were measured,
then compared with the rest (laboratory) values. For the objects observed at
two different epochs, measurements were obtained for each spectrum. The measured
heliocentric radial velocities, and their estimated standard deviations, are listed in
Table~\ref{tab:obs2}.

Measurements of radial velocities have been previously obtained by a number of authors for
several stars in our sample. However, most of them were based on low- or
medium-resolution spectroscopy (e.g., Bond 1980; Norris, Bessell, \& Pickles
1985). We choose to compare our measurements with only the results of
high-dispersion spectroscopy from recent studies. The comparisons are given in
Table 3. No significant variation in radial velocity is found for most objects.
HD~4306 and HD~186478 show changes of radial velocity of about 6 km~s$^{-1}$ and
2 km~s$^{-1}$, respectively, suggesting that both stars may be members of binary
systems. Indeed, Carney et al.(2003) have obtained an orbital
solution for HD~186478, showing it to be a long-period binary ($P \approx 550$
days) with a low amplitude, on the order of 3 km~s$^{-1}$. No definitive
conclusion can yet be achieved for HD~4306, hence further radial velocity
monitoring is required to confirm its possible binarity. We note that
HD~186478 is a moderately r-process-enhanced star, with [Eu/Fe] $\approx +0.5$
(Johnson \& Bolte 2001). Preston \& Sneden (2001) investigated the
variations of radial velocity for the extremely r-process-enhanced star
CS~22892-052 in detail. However, there is still no clear evidence of binarity,
as the suspected amplitude of the variation is quite small. We did not find
evidence of binarity for any of our other r-process-enhanced stars.

\section{Summary}

We have obtained high-resolution, high-S/N ratio, spectra for 22 very metal-poor
stars with Subaru/HDS, taken during the commissioning phase of this instrument.
These stars were selected so as to include as many objects with known (or
suspected) enhancement of r-process elements as possible. In this paper we have
reported the measurements of equivalent widths for isolated absorption lines in
the reduced spectra, and also precision radial velocities (in some cases, at
several epochs), for each star. Comparisons of our measured equivalent widths
with previous work demonstrates that there exists no systematic differences, except
in the cases of stronger lines in a few objects for which the S/N ratios of the
previous work was rather low. In the following papers of this series (Paper II,
others in preparation), the results of the detailed abundance analysis for these
data will be presented.

\acknowledgments

We thank all of staff members of the Subaru telescope, NAOJ, for their help
during the observations. T.C.B. acknowledges partial support from grants AST
00-98508 and AST 00-98548 awarded by the U.S. National Science Foundation. Most
of the data reduction was carried out at the Astronomical Data Analysis Center
(ADAC) of the National Astronomical Observatory, Japan.

\clearpage

\begin{deluxetable}{clcccccc} 
\tablewidth{0pt}
\tablecaption{PROGRAM STARS AND OBSERVATION LOG \label{tab:obs}}
\startdata
\tableline
\tableline
\multicolumn{1}{c}{No} & Star & $V$ & $B-V$ & $R$ & S/N$^{\rm a}$ & S/N$^{\rm b}$ & Exp.(sec)\\
\tableline
1&HD~4306 & 9.08 & 0.63 & 90000 & 272 & 497 & 1800\\
2&HD~6268 & 8.10 & 0.79 & 90000 & 158 & 288 & 1800\\
3&HD~88609 & 8.59 & 0.93 & 90000 & 62 & 113 & 2110\\
4&HD~110184 & 8.31 & 1.17 & 90000 & 221 & 403 & 900\\
5&HD~115444 & 8.97 & 0.78 & 90000 & 255 & 466 & 3900\\
6&HD~122563 & 6.20 & 0.91 & 90000 & 374 & 683 & 1200\\
7&HD~126587 & 9.15 & 0.73 & 90000 & 187 & 341 & 4500\\
8&HD~140283 & 7.21 & 0.49 & 90000 & 458 & 836 & 3600\\
9&HD~186478 & 9.18 & 0.90 & 100000 & 158 & 274 & 2400\\
10&BS~16082--129 & 13.55 & 0.67 & 50000 & 55 & 135 & 5400\\
11&BS~16085--050 & 12.15 & 0.74 & 50000 & 100 & 245 & 5100\\
12&BS~16469--075 & 13.42 & 0.77 & 50000 & 59 & 145 & 5400\\
13&BS~16920--017 & 13.88 & 0.76 & 50000 & 41 & 100 & 5400\\
14&BS~16928--053 & 13.47 & 0.85 & 50000 & 49 & 120 & 5400\\
15&BS~16929--005 & 13.61 & 0.62 & 50000 & 56 & 137 & 5400\\
16&BS~17583--100 & 12.37 & 0.51 & 50000 & 79 & 194 & 3600\\
17&CS~22169--035 & 12.88 & 0.92 & 50000 & 49 & 120 & 5400\\
18&CS~22183--031 & 13.62 & 0.65 & 50000 & 47 & 115 & 9751\\
19&CS~22892--052 & 13.18 & 0.78 & 90000 & 60 & 147 & 7200\\
20&CS~22952--015 & 13.27 & 0.78 & 50000 & 59 & 145 & 9000\\
21&CS~30306--132 & 12.81 & 0.80 & 50000 & 85 & 208 & 4493\\
22&CS~31082--001 & 11.67 & 0.77 & 50000 & 100 & 122 & 1200\\
\tableline
\enddata
~ \\
$^{\rm a}$ S/N ratio per pixel at 4000~{\AA}.  

$^{\rm b}$ S/N ratio per resolution element at 4000~{\AA}. 
\end{deluxetable}

\clearpage

\begin{deluxetable}{clccccccl} 
\tablewidth{0pt}
\tablecaption{COORDINATES AND RADIAL VELOCITY \label{tab:obs2}}
\startdata
\tableline
\tableline
\multicolumn{1}{c}{No} & Star & R.A. (J2000.0) & Dec. (J2000.0)&
 Obs.Date& Vr $^{\rm a}$\\
\tableline
1&HD~4306 & 00 45 27.2 & $-$09 32 40 & 19 Aug, 2000 & --69.69$\pm$0.29\\
2&HD~6268 & 01 03 18.2 & $-$27 52 50 & 18 Aug, 2000 & 39.20$\pm$0.27\\
3&HD~88609 & 10 14 29.0 & +53 33 39 & 11 nov, 2000 & --37.28$\pm$0.43\\
4&HD~110184 & 12 40 14.1 & +08 31 38 & 29 Jan, 2001 & 138.89$\pm$0.27\\
5&HD~115444 & 13 16 42.5 & +36 22 53 & 4 July, 2000 & --27.30$\pm$0.34\\
5&HD~115444 & & & 28 Jan, 2001 & --27.58$\pm$0.25\\
6&HD~122563 & 14 02 31.9 & +09 41 10 & 4 July, 2000 & --27.20$\pm$0.33\\
6&HD~122563 & & & 29 Jan, 2001 & --26.52$\pm$0.34\\
7&HD~126587 & 14 27 00.4 & $-$22 14 39 & 27 Jan, 2001 & 148.72$\pm$0.72\\
7&HD~126587 & & & 31 Jan, 2001 & 149.10$\pm$0.24\\
8&HD~140283 & 15 43 03.1 & $-$10 56 01 & 4 July, 2000 & --171.17$\pm$0.29\\
8&HD~140283 & & & 17 Aug, 2000 & --170.23$\pm$0.19\\
9&HD~186478 & 19 45 14.1 & $-$17 29 27 & 20 Aug, 2000 & 30.52$\pm$0.31\\
10&BS~16082--129 & 13 47 11.5 & +28 57 46 & 30 Jan, 2001 & --92.16$\pm$0.30\\
11&BS~16085--050 & 12 37 46.7 & +19 22 44 & 31 Jan, 2001 & --75.06$\pm$0.28\\
12&BS~16469--075 & 10 15 10.1 & +42 53 19 & 28 Jan, 2001 & 332.88$\pm$0.74\\
13&BS~16920--017 & 12 07 17.1 & +41 39 35 & 27 Jan, 2001 & --206.53$\pm$0.84\\
14&BS~16928--053 & 12 22 28.1 & +34 11 24 & 28 Jan, 2001 & --81.00$\pm$0.35\\
15&BS~16929--005 & 13 03 29.4 & +33 51 06 & 30 Jan, 2001 & --51.29$\pm$0.43\\
16&BS~17583--100 & 21 42 27.8 & +26 40 34 & 19 Aug, 2000 & --107.96$\pm$0.32\\
16&BS~17583--100 &  &  & 20 Aug, 2000 & --108.76$\pm$0.42\\
17&CS~22169--035 & 04 12 13.9 & $-$12 05 05 & 11 Nov, 2000 & 17.72$\pm$0.73\\
18&CS~22183--031 & 01 09 04.9 & $-$04 43 25 & 10 Nov, 2000 & 11.67$\pm$0.67\\
18&CS~22183--031 & & & 11 Nov, 2000 & 11.97$\pm$0.56\\
19&CS~22892--052 & 22 17 01.5 & $-$16 39 26 & 22 July, 2001 & 12.72$\pm$0.49\\
20&CS~22952--015 & 23 37 28.6 & $-$05 47 56 & 11 Nov, 2000 & --20.07$\pm$0.71\\
21&CS~30306--132 & 15 14 18.6 & +07 27 02 & 26 July, 2001 & 109.01$\pm$0.31\\
22&CS~31082--001 & 01 29 31.2 & $-$16 00 48 & 30 July, 2001 & 138.91$\pm$0.30\\
\tableline
\enddata
~ \\
$^{\rm a}$ Heliocentric radial velocity (km~s$^{-1}$).  
\end{deluxetable}

\clearpage

\begin{deluxetable}{cp{6mm}p{3mm}p{9mm}p{5mm}p{5mm}p{5mm}p{5mm}p{5mm}p{5mm}p{5mm}p{5mm}p{5mm}p{5mm}p{5mm}p{5mm}p{5mm}p{5mm}p{5mm}p{5mm}p{5mm}p{5mm}p{5mm}p{5mm}p{5mm}p{5mm}}
\rotate
\tabletypesize
\scriptsize
\tablewidth{0pt}
\tablecaption{EQUIVALENT WIDTHS FOR PROGRAM STARS: FIRST SAMPLE \label{tab:ew1}}
\startdata
\tableline
\tableline
\multicolumn{1}{c}{Wavelength} & Species & \multicolumn{1}{c}{L.E.P.} & \multicolumn{1}{c}{$\log gf$} & \multicolumn{9}{c}{Equivalent width (m{\AA})$^{\rm a}$}  \\
  \cline{5-26} 
\multicolumn{1}{c}{(\AA)} &  & \multicolumn{1}{c}{(eV)} & \multicolumn{1}{c}{ } & \multicolumn{1}{c}{1} & \multicolumn{1}{c}{2} & \multicolumn{1}{c}{3} & \multicolumn{1}{c}{4} & \multicolumn{1}{c}{5} & \multicolumn{1}{c}{6} & \multicolumn{1}{c}{7} & \multicolumn{1}{c}{8} & \multicolumn{1}{c}{9} & \multicolumn{1}{c}{10}& \multicolumn{1}{c}{11} & \multicolumn{1}{c}{12} & \multicolumn{1}{c}{13} & \multicolumn{1}{c}{14}& \multicolumn{1}{c}{15}& \multicolumn{1}{c}{16} & \multicolumn{1}{c}{17} & \multicolumn{1}{c}{18} & \multicolumn{1}{c}{19}& \multicolumn{1}{c}{20} & \multicolumn{1}{c}{21} & \multicolumn{1}{c}{22}\\
\tableline

3829.35 &  Mg I  &2.71 &  --0.480  &157.6&198.5&156.4&    &    &    &    &125.3&      &131.5&158&131.2&110.3&151.6&109.4&125.2&154&131.5&140.6&96.5&172.5& \\
3832.31 &  Mg I  &2.71 &    0.145  &190.1&213.5&160.4&263.1&181.6&230&174.9&153.2&231.7&143.4&177&137.9&103.6&142.5&120&142.2&124.4&122.6&158.7&103.9&208.8&191\\
3838.30 &  Mg I  &2.72 &    0.414  &237.7&261.9&194.9&    &205.4&262&202&173.5&288.8&166.2&198&163.7&137.1&171&127.9&164.7&156.6&143.7&176.2&120.5&265.3&     \\
4571.10 &  Mg I  &0.00 &  --5.569  &45.5&77.4&76.5&132.6&51.7&85&43.8&7.7&95.7&32.2&42.2&24.1&17.5&49&15.4&10.9&49.7&21.6&27.4&17.1&59.2&55.3\\
4703.00 &  Mg I  &4.35 &  --0.377  &    &    &    &    &54.6&72&52.6&41&90.7&45&56.5&    &    &    &    &39.9&    &38.2&41.5&23.7&74.3&61.9\\
5172.70 &  Mg I  &2.71 &  --0.381  &    &    &    &    &    &209&168.4&    &      &150.4&176.8&    &131.6&    &    &    &    &    &154.5&    &193&187.3\\
5183.62 &  Mg I  &2.72 &  --0.158  &    &    &    &    &    &232&187.1&    &      &169.6&192.4&    &149.7&    &    &    &    &    &177.4&    &231.7&     \\
3961.53 &  Al I  &0.01 &  --0.336  &106.9&139.5&118.1&174.3&112.4&146&104.6&68.4&145.3&88&105&85.5&85.1&97.5&64.2&67.1&101.6&81.4&107&93.2&119.1&110.2\\
4102.94 &  Si I  &1.91 &  --3.100  &63&91.4&76.6&114.5&57.4&82&60.5&19.4&89&46.8&73.2&40&22.9&56.1&18.2&15.5&68.3&40.1&44.6&45.6&68.5&66.9\\
4226.73 &  Ca I  &0.00 &    0.240  &    &    &    &    &    &    &    &    &      &    &    &    &    &    &    &134.9&    &    &  syn  &    &    &     \\
4283.01 &  Ca I  &1.89 &  --0.220  &    &    &    &    &46.5&65.6&52.2&29.6&84&    &39.5&35.5&12.8&44.2&30&28&    &16.5&    &18.3&    &     \\
4318.66 &  Ca I  &1.90 &  --0.210  &47.4&64.1&43.9&81.4&40.9&55.4&40.6&27.2&69.9&37.2&36.8&24.9&25.2&40.1&18.8&26.7&    &19.3&32&26.8&54.4&48.7\\
4425.44 &  Ca I  &1.88 &  --0.358  &42.7&55&39.9&71.6&34.4&49.1&35.7&22.2&63.5&33.9&31.4&29.2&11.2&41.8&20.7&22.4&    &19&35.2&    &52&     \\
4454.79 &  Ca I  &1.90 &    0.260  &70.3&95.4&69.5&    &64.9&79.9&64.6&48.5&97.1&57.9&60.7&42.6&30.8&61.7&35.1&43.9&    &49.9&    &    &81&     \\
4455.89 &  Ca I  &1.90 &  --0.510  &    &    &35.4&    &    &    &    &    &      &24.2&    &    &    &    &    &17.5&    &    &25&    &    &     \\
4400.40 &  Sc II  &0.60 &  --0.540  &56.2&86.6&67&    &58.2&78.3&    &17.5&79.6&    &    &    &    &    &    &13.5&    &30.5&    &32.3&    &     \\
4415.56 &  Sc II  &0.59 &  --0.670  &48.9&78.1&61.7&103&52.2&76&47.7&14.7&80.9&37.4&51.4&29.1&29.2&41.1&3.7&13.1&46.6&22.8&41.4&23.5&50.7&     \\
5031.02 &  Sc II  &1.36 &  --0.400  &20.3&46&30.2&64.7&22.3&42&18.5&4.8&47&17.3&27.1&12.3&8.4&21.9&    &    &20.3&    &15.3&    &26.2&21.8\\
3998.64 &  Ti I  &0.05 &  --0.056  &54.8&    &    &108.8&56.5&70.2&52.3&24.2&81.5&43.3&38.3&28&    &53&26.2&23.3&    &32.8&37.7&33.2&59.4&59.2\\
4533.25 &  Ti I  &0.85 &   0.476  &38.6&    &    &84.6&40.9&52.2&35.5&14.4&63.4&29.1&24.8&20.6&30&36&    &    &    &54.3&25.3&23&47.5&104.4\\
4534.78 &  Ti I  &0.84 &   0.280  &    &    &36.5&71.8&31.1&42.6&95.5&    &54.7&23.4&17.6&16.1&19.3&30.9&14.1&10.6&36&    &21.6&    &38.1&32\\
4535.58 &  Ti I  &0.83 &   0.130  &    &    &    &    &24.2&35.6&21.2&    &49&18.8&12.4&    &14.7&29.9&10.5&    &    &    &15.1&    &30.7&27.2\\
4981.74 &  Ti I  &0.85 &   0.504  &44.5&66.7&51&97.5&47.2&61.6&39.8&19.1&73.3&36.9&31&25&39.2&    &16.7&19.4&36.7&22.5&31.5&15.9&54.7&48.9\\
4991.07 &  Ti I  &0.84 &   0.380  &40.7&    &    &    &42.1&57.8&37.7&16.4&71&34.1&24.5&22.4&17.8&39.7&16&    &36.6&16.1&29.1&    &49.5&44.8\\
4999.51 &  Ti I  &0.83 &   0.250  &35.8&57.3&40.4&86.1&36.3&49.1&31.6&12&64.3&23.2&21.7&16.4&25.3&31&13.6&15.1&    &16.7&    &12.8&42.2&47.3\\
5039.96 &  Ti I  &0.02 &  --1.130  &17.3&    &22.7&72.1&18.9&30.2&15.9&3.3&45.3&13.8&10.5&10.5&12.8&14.9&    &    &    &    &9.7&    &20&20.6\\
5064.66 &  Ti I  &0.05 &  --0.991  &20.5&    &29.9&78.9&22.5&35.9&18&5&49.3&10.8&11.1&11.3&    &22.3&    &    &    &    &12.4&    &27.9&     \\
5173.75 &  Ti I  &0.00 &  --1.118  &    &    &    &76.3&    &34&16.6&    &      &    &8.8&8.1&14.8&20.2&    &    &    &    &    &    &22.1&     \\
5192.98 &  Ti I  &0.02 &  --1.006  &    &    &    &    &    &39.2&19.7&    &      &17.9&11.5&10.7&15.5&18.7&7&    &    &    &    &    &28.8&     \\
4028.35 &  Ti II  &1.89 &  --1.000  &38.4&    &    &82.2&42.1&52.5&35.4&13.4&70.9&27.7&25.2&21.8&26.8&33.5&12&11.3&    &16.8&    &22.4&45.8&     \\
4337.93 &  Ti II  &1.08 &  --1.130  &84.5&119&80.4&145.1&92.8&    &78&40.9&110.1&    &    &    &70.2&    &    &37.4&69.6&47.5&    &50&81&     \\
4394.07 &  Ti II  &1.22 &  --1.590  &42.3&70&55&    &48.1&    &    &11.5&74.9&    &    &    &    &    &    &12.2&    &    &    &    &    &     \\
4395.85 &  Ti II  &1.24 &  --2.170  &30.6&57.8&122.6&    &36.2&46.6&    &6.7&58.1&    &    &    &    &    &    &    &103.4&64.8&    &    &    &     \\
4399.78 &  Ti II  &1.24 &  --1.270  &74.5&96.6&81.4&    &75.4&89&    &30.5&102.5&    &    &    &    &    &    &29&53.7&42.4&    &35.9&    &     \\
4417.72 &  Ti II  &1.17 &  --1.430  &73.9&105&88.6&129.5&81.1&95.1&70.6&32.3&104.1&62.2&60.2&52.6&63.4&69.4&29.4&28.8&67.6&34.4&65.7&41.4&86.5&     \\
4443.81 &  Ti II  &1.08 &  --0.700  &96.8&128.2&116&160.1&104&119.3&95.7&60.8&126.7&85.4&90.2&76.3&87.7&100.8&55.3&57.7&91.6&61.1&84.7&73.3&100.6&101.9\\
4450.49 &  Ti II  &1.08 &  --1.450  &64.7&96.6&75.6&124.7&71.5&87.4&63.8&23.8&92.5&    &54.3&43.2&58.1&65.8&23.8&21.6&52.3&34.7&50.9&33.9&68.5&68.9\\
4464.46 &  Ti II  &1.16 &  --2.080  &44.8&    &    &    &50.2&67&42.9&11.7&76.5&35.9&32.6&22.7&38.5&45.8&18.3&12.7&    &    &30.5&    &50.5&47.1\\
4468.50 &  Ti II  &1.13 &  --0.600  &98.4&132.5&112.8&163.8&107.1&121.4&98.1&63.1&130.6&83.7&90.6&78.3&89.3&101&59.6&60.3&88.6&57.8&93.3&70.9&102.4&     \\
4470.86 &  Ti II  &1.17 &  --2.280  &28.1&56.8&38.1&74.9&34.1&48.6&26.6&5.4&59.3&22.4&18.9&16.6&24.5&27.3&    &    &    &    &    &    &34.6&     \\
4501.28 &  Ti II  &1.12 &  --0.750  &93.9&    &109.8&156.1&101.7&117.2&93.1&57.2&125.3&82.7&86.3&70.7&83&95&55.6&54.2&92.5&60.7&83.2&69.9&97.9&98.5\\
4571.98 &  Ti II  &1.57 &  --0.530  &85.8&    &    &154.6&92.6&109.9&84.5&52.6&117&74&80.5&64.4&76.8&87.9&44.7&49.2&85&49.1&77.2&62.1&92.2&92.3\\
4589.95 &  Ti II  &1.24 &  --1.790  &48.4&    &61.5&102.4&54.3&71.7&46.4&13.9&78.9&40&34.3&26.2&41.6&48.5&10.2&12.5&47.9&18.4&34&    &52.4&55.4\\
4865.62 &  Ti II  &1.12 &  --2.610  &11.3&29.7&20.9&51.8&15.1&25.2&11.5&2&32.3&12&9.8&    &    &22&    &    &    &    &11.1&    &14.7&     \\
5129.16 &  Ti II  &1.89 &  --1.390  &    &    &    &72.6&    &45.1&22.8&    &      &23.2&17.2&12.9&20.6&24.9&    &    &    &    &18.4&    &31.6&     \\
5185.91 &  Ti II  &1.89 &  --1.350  &    &    &    &65.6&    &37.1&19&    &      &16.5&15.4&11.1&17.9&22.6&    &    &    &    &11.8&    &27.1&22\\
5188.70 &  Ti II  &1.58 &  --1.210  &    &    &    &    &    &90&59.8&    &      &    &    &34&49&    &19.2&    &    &    &    &    &    &     \\
4379.24 &  V I  &0.30 & 0.565  &14.2&28.9&    &    &14&28&    &4.9&38.6&    &    &    &    &    &    &    &    &    &    &11.6&    &     \\
4389.99 &  V I  &0.28 &   0.235 &    &    &    &    &8.7&21.7&    &    &      &    &    &    &    &    &    &    &    &    &    &    &    &     \\
3951.96 &  V II  &1.48 &  --0.784  &17.7&44.6&34.9&    &21.6&35.4&17.4&5.2&42.5&36.1&18.2&11.6&    &16.2&    &6.2&    &17.1&15.7&    &23.3&29.4\\
4005.71 &  V II  &1.82 &  --0.522  &18.1&46&38.6&    &19.3&36.3&18.4&5.2&43.5&16.9&17.6&7.1&17.6&21.1&    &    &    &    &    &    &23.2&33.5\\
4254.35 &  Cr I  &0.00 &  --0.114  &    &    &    &    &    &113.7&88&66.3&      &84.2&83.7&70.7&82.9&95.6&52.4&60.1&    &    &87.4&    &96.8&95.7\\
4274.81 &  Cr I  &0.00 &  --0.231  &    &    &    &    &    &112.8&88.1&63.5&      &82.2&79.4&69.1&79.6&92.5&52.9&53.6&    &51.3&    &    &    &96.7\\
4289.73 &  Cr I  &0.00 &  --0.361  &85.9&109.8&92.1&135.1&86&109.1&82&57.2&      &76.8&72.6&67.2&72.9&88.3&47.6&49.6&88.1&53.1&    &66.4&96.1&     \\
4554.99 &  Cr II  &4.07 &  --1.380  &    &    &    &    &2.8&4.8&4.8&1.5&      &    &    &    &    &    &    &    &    &    &    &    &    &     \\
4558.65 &  Cr II  &4.07 &  --0.660  &10.2&30.2&13.4&    &11.1&20.7&20.7&8.7&28.2&9.8&11.2&    &8.9&12.2&    &    &    &    &    &    &17.9&     \\
4588.20 &  Cr II  &4.07 &  --0.630  &8.4&19.8&6.8&23.5&7.1&13.3&13.3&5&18.6&5.2&5&    &4.1&7.9&    &    &    &    &    &    &9&     \\
4030.76 &  Mn I  &0.00 &  --0.470  &91&138.2&105.5&    &97.2&140.8&97.2&66.8&138.8&84.8&106.1&75.5&98.7&97.1&39.5&50.4&130.3&65.1&93.6&91.6&109.8&109\\
4033.07 &  Mn I  &0.00 &  --0.618  &77.4&124.4&95.8&179.3&83.6&125.2&83.6&55.2&124.9&75.1&91.6&61.3&86.7&91.9&35.6&38.9&112.2&46.3&92.2&73.2&97.8&96.2\\
4034.49 &  Mn I  &0.00 &  --0.811  &79.1&105.8&89.3&164.8&75.2&119.8&75.2&45.9&121.3&77.1&80.9&57.7&80.8&86.7&    &33&    &36.4&    &57.4&108.1&91.8\\
4041.37 &  Mn I  &2.11 &    0.285  &16.6&45.7&37.2&67.7&15.7&43.9&15.7&13.3&47.1&23.2&26.6&7.7&31.8&26.6&    &    &    &    &18.7&    &30.8&19.8\\
4754.04 &  Mn I  &2.28 &  --0.086  &6.3&18.1&11.6&40.6&5.9&19.9&5.9&5.2&21&6.5&9.7&    &13.5&13.2&    &    &    &    &5.5&    &12.3&7.3\\
4823.51 &  Mn I  &2.32 &    0.144  &8.8&26.2&12.8&    &8.6&25.7&8.6&7.3&28.4&11.3&14.9&    &13.3&14.3&    &    &    &    &    &    &20.1&     \\
3763.80 &  Fe I  &0.99 &  --0.221  &139&209.9&    &    &152.4&171.7&138.7&107.8&217.1&113.9&131.3&116.6&98&137.2&    &103&    &102.5&134.5&53.9&155&140.4\\
3767.20 &  Fe I  &1.01 &  --0.382  &125.9&175.7&    &    &137.6&151.7&127.3&96.9&180.3&106.1&120.7&110.8&90.8&128.9&83.8&86.4&    &84.4&127.6&109.2&137.8&     \\
3787.89 &  Fe I  &1.01 &  --0.838  &112&148.2&116&222.8&131.1&129.7&110.3&84.5&145.9&89.4&108.4&100.5&85.3&115.7&72&80.9&130.3&63.7&109&51.2&114.8&109\\
3805.35 &  Fe I  &3.30 &    0.313  &46.8&78.8&    &81.6&58.1&59.3&48.9&40.4&76.9&52.1&42.3&42.7&    &47.8&19.2&34.3&    &31.2&65.3&47.4&63.6&     \\
3815.85 &  Fe I  &1.49 &    0.237  &137.2&191.1&    &    &150.9&165.6&138.1&114.6&195.1&119.6&131.5&121.5&96.9&138.2&91.8&109&    &111.1&138.7&102.5&156.6&164\\
3820.44 &  Fe I  &0.86 &    0.157  &188.2&287.4&252.6&    &210.3&279.7&193.3&148.5&331.7&168.2&172.2&159.9&149.9&210.2&122&134.8&    &143.3&183.2&149.6&234&210\\
3825.89 &  Fe I  &0.92 &  --0.024  &165.1&234.8&198.3&    &182.3&222.2&165.8&129.4&272.9&151.6&152.4&134.5&131&180&107&119&    &125.5&171.2&135.5&193.3&     \\
3827.83 &  Fe I  &1.56 &    0.094  &124.3&177.5&117&    &136&159&124.5&102&185.9&112.5&122.1&97.8&107.9&137.8&84.4&    &    &89.7&127.6&101.5&140&     \\
3840.45 &  Fe I  &0.99 &  --0.497  &123.9&171.5&138.8&    &136.6&154.4&126.7&96.1&182.8&109.9&121.7&111.7&92.1&130&82.2&91.7&    &88.9&121.1&109.8&139&129.4\\
3849.98 &  Fe I  &1.01 &  --0.863  &110.6&149.7&    &    &125.2&142.2&111.9&87.3&156.4&103.7&111.7&101.2&99.5&124.7&82.3&80.3&    &78.6&117.3&105.2&119.1&119.9\\
3856.38 &  Fe I  &0.05 &  --1.280  &137.6&180.7&160.6&    &155.8&173.6&140.3&106.1&189.3&117.6&139.6&117.5&123.1&148.7&96&98&155.9&101.5&138.1&138.9&146.4&149\\
3859.92 &  Fe I  &0.00 &  --0.698  &184&234.7&203.7&    &190.7&238.2&185.3&137.5&292.4&151.9&168.5&156&142.5&195.6&116.1&126.6&182.3&121&179&144.1&215.2&191.1\\
3865.53 &  Fe I  &1.01 &  --0.950  &111.2&144.8&115.9&194.6&119.6&136.1&111.7&84.1&148.2&100.7&110.3&100.9&90.4&117.7&72.5&78.7&128.6&82.2&111.1&102&115.8&118.7\\
3886.29 &  Fe I  &0.05 &  --1.055  &115.4&208.5&161.7&    &165.3&184.7&156&109&221.3&127.6&141.4&144.7&125.8&150.4&103.5&105.5&174.1&104.8&193.4&132.8&179.6&173.3\\
3899.72 &  Fe I  &0.09 &  --1.515  &127.7&165.9&150.1&    &140.5&162.5&129.6&96.7&168.2&121&133.8&122.7&118&140.5&93.8&93.3&158.4&91&130.7&133.3&132.7&139.3\\
3902.96 &  Fe I  &1.56 &  --0.442  &103.8&138.9&118.5&    &112.1&131.4&105.5&82.2&123.5&98.2&103.5&89.1&92.4&111.8&68.8&    &    &76.3&105.9&11.3&111.3&108.1\\
3920.27 &  Fe I  &0.12 &  --1.734  &121.8&161.7&142.5&    &137&157.7&123.1&89.3&161.7&119.8&122.6&115.9&114.1&137.1&88.4&84.6&    &99.5&125.8&118.2&121.8&126.2\\
3922.92 &  Fe I  &0.05 &  --1.626  &128.9&171.7&148.1&253.8&142.5&168.4&132&96.1&173.5&124.2&132.4&120.6&114.2&148.2&95.1&91.4&164.3&93.6&128&130.3&135.7&133.5\\
3949.96 &  Fe I  &2.18 &  --1.251  &45.5&68.5&61&95&47.7&66.9&44.9&28.3&73.8&37.2&36.1&30.8&26.8&48.1&16.5&23.2&59.4&36.9&42.9&30.1&55.9&50.7\\
4005.25 &  Fe I  &1.56 &  --0.583  &101.5&142.7&110.1&    &110.4&131.7&103.1&78.8&142.4&98.8&99.4&87&85&112&67.2&71.5&    &81.8&100.4&102.5&104.7&107.5\\
4063.61 &  Fe I  &1.56 &    0.062  &128.4&186.3&147.7&    &139.4&166.3&129.6&106.7&179.9&118.6&127.5&115.9&110.3&137.6&88.3&96.5&    &110.1&135.7&123.7&144.9&     \\
4071.75 &  Fe I  &1.61 &  --0.008  &120.6&168.7&150&    &130.1&153.7&120.8&    &171.2&109.6&117.3&107.4&72.6&115.7&80.9&91.7&    &76.9&117.3&113.8&130.9&128.7\\
4076.64 &  Fe I  &3.21 &  --0.528  &26.7&    &25.8&    &29.4&45.1&27.2&18.5&59.8&    &21.9&    &    &    &16&15.5&    &    &34.6&     &39.4&     \\
4114.45 &  Fe I  &2.83 &  --1.303  &12.8&33.3&16.8&52.8&14.3&28.8&13&8.7&38.9&19.6&10.7&    &    &16.7&    &    &    &    &10.2&17.4&18.8&16.6\\
4132.91 &  Fe I  &2.85 &  --1.005  &23.1&47.2&119.5&184.4&22.7&42.2&102.7&15.3&143.8&90.2&    &87.2&    &24.7&    &17.3&    &67.3&101.6&95.4&33.8&     \\
4134.69 &  Fe I  &2.83 &  --0.649  &36.3&66&48&    &43.8&58.3&38&28.2&68.6&37.1&    &21.4&    &37.8&15.7&21.3&    &18.5&34.6&33.8&53.6&48.9\\
4143.88 &  Fe I  &1.56 &  --0.511  &104&134.7&123.9&    &113.2&134.5&104.8&83.7&135.3&100.1&105.5&94.4&102.3&117.5&75.9&81.9&    &72.5&104.2&101.3&111.1&50.4\\
4147.68 &  Fe I  &1.49 &  --2.071  &47.5&82.1&64&112.5&57.5&78.7&49.5&22.5&85.5&46.6&40&35.3&    &59.9&18&15.2&    &15.7&49.2&35.6&58.7&57.9\\
4154.51 &  Fe I  &2.83 &  --0.688  &38.1&49.4&36.9&    &32.1&52.4&23&21.3&64.7&27.2&24&    &    &36.2&    &21.5&    &    &    &    &51.9&     \\
4156.81 &  Fe I  &2.83 &  --0.808  &39.6&63&37.5&    &34.5&59.3&33.6&22.5&69.2&29&27.1&17.9&    &37&    &    &    &    &28.5&    &46.1&     \\
4157.79 &  Fe I  &3.42 &  --0.403  &27.8&38.8&24&58.2&22.2&    &22.7&13.9&57.7&19.7&15.6&17&    &    &    &    &    &    &29.2&    &35.6&     \\
4174.92 &  Fe I  &0.92 &  --2.938  &41.9&76.2&70.2&130.4&58.1&78.6&44.7&14.3&84.6&    &31.9&25.7&30.9&    &    &    &66.9&    &45&33.8&57.3&53.8\\
4175.64 &  Fe I  &2.85 &  --0.827  &30.1&66.2&41.8&78.9&42.4&52.5&31.1&20.9&64.5&    &20.9&    &    &    &14.8&17.7&    &    &41.3&20.1&45.6&     \\
4181.76 &  Fe I  &2.83 &  --0.371  &53.2&79.5&57.2&    &    &69.8&50.5&38.7&84.2&47.1&41.2&38.8&    &53.5&25.2&    &    &    &56.5&    &59.6&57.2\\
4182.39 &  Fe I  &3.02 &  --1.180  &8.6&25.9&10.2&40.9&    &18.9&10.2&5.7&27.5&    &    &    &    &    &    &    &    &    &    &    &15.6&     \\
4187.05 &  Fe I  &2.45 &  --0.514  &64.7&91.4&77.4&123.5&82.7&    &64&51.7&94.1&53.6&55.9&51.5&44.4&74.2&31.8&38.7&86.7&35.3&61.8&55.4&74.1&71.8\\
4187.81 &  Fe I  &2.42 &  --0.510  &66.5&97.9&77.1&    &70.4&    &69&57.8&103.1&63.7&59.6&55&54.6&73.5&38.3&40.6&    &31.7&69.6&57.8&80.6&75.7\\
4191.44 &  Fe I  &2.47 &  --0.666  &54.7&90.3&61.6&    &59.2&79.5&56.8&    &84.8&51.7&50.8&41.3&41.8&62.1&22.7&32.9&    &    &55.3&    &69.2&68.8\\
4195.34 &  Fe I  &3.33 &  --0.492  &30.5&39.3&25.5&73.7&30.3&43&27.7&16.6&56.4&28.7&17.3&17.3&    &28.3&    &16&    &    &    &    &41.4&32.3\\
4199.10 &  Fe I  &3.05 &    0.156 &61&85.3&74.3&    &63.7&80.9&60.4&48.3&90.2&56.2&53.6&47.5&38&63.7&32.2&43.1&78.1&28.3&72.7&51.3&70.6&74.2\\
4202.04 &  Fe I  &1.49 &  --0.689  &102.3&137.5&113.1&    &110&132.1&104.6&80.2&126.2&99.4&101.3&90.3&94.2&112.2&72.4&76.3&140.3&73.7&100.9&    &106.7&115\\
4222.22 &  Fe I  &2.45 &  --0.914  &45.6&74.4&    &101.9&48.6&73.4&47.3&    &78.6&46.6&40.3&32.7&29.9&58.1&20.1&    &66.2&25.9&40.8&30.1&59.2&60.1\\
4227.44 &  Fe I  &3.33 &    0.266  &    &93.7&88.3&    &64&96&59.6&42.8&      &53.8&49.4&39.7&40.2&67.7&28.8&    &    &    &54&38.4&76.9&68.8\\
4233.61 &  Fe I  &2.48 &  --0.579  &    &    &    &119.7&    &85.9&60.5&43.5&      &57.2&54.6&45.3&58.9&66.7&29.5&    &    &    &58.3&    &73.4&68.3\\
4250.13 &  Fe I  &2.47 &  --0.380  &    &    &    &123.3&    &96.3&70.1&53.8&      &61.7&62.3&53.8&46.7&74.7&34.5&    &    &    &63.8&    &78.2&83.5\\
4250.80 &  Fe I  &1.56 &  --0.713  &    &    &    &166.2&    &129.9&100.6&77.8&      &98.8&98.1&89.8&87.6&108.5&62&    &    &    &96.1&    &105&109.8\\
4260.49 &  Fe I  &2.40 &    0.077  &    &    &    &    &    &122.6&95&77.9&      &88.2&91&79.1&79.9&98.3&55.7&    &    &    &88.5&    &100.7&99.4\\
4271.16 &  Fe I  &2.45 &  --0.337  &    &    &    &    &    &106.9&82.8&55.4&      &    &70.5&64.8&118.9&84.3&54.8&    &    &    &    &    &101.2&     \\
4325.77 &  Fe I  &1.61 &    0.006  &125.6&173.9&150.7&    &134&163.3&130.9&101.3&180.7&    &124&    &123.8&140.8&89&97.4&    &97.5&    &123.7&146.3&     \\
4337.06 &  Fe I  &1.56 &  --1.704  &62.1&94.1&74.1&136.2&69.4&88.9&65.5&32.8&93.5&    &    &49.6&    &    &    &32.7&    &40.1&    &44.6&76.3&     \\
4383.56 &  Fe I  &1.49 &    0.208 &143.6&195.1&175.6&    &151.5&188.2&    &122.4&199.3&    &    &    &    &    &    &111.8&178.7&    &    &136.9&    &     \\
4404.76 &  Fe I  &1.56 &  --0.147  &123.3&168.1&146.1&    &130&    &    &101.2&171.2&    &    &    &    &    &    &94.5&161.5&90.3&    &117.5&    &     \\
4415.14 &  Fe I  &1.61 &  --0.621  &104&126.4&126.5&    &109&134.8&108&82.2&136.7&    &102.2&94&100.3&    &69.6&76.7&135.6&82.3&111.7&101.7&    &     \\
4430.62 &  Fe I  &2.22 &  --1.728  &26.7&55.7&39.1&86.5&31.3&54.7&27&13.8&61.8&28&    &19.2&    &39.1&    &    &    &    &26.9&22.4&40.1&     \\
4442.35 &  Fe I  &2.20 &  --1.228  &49.3&82.2&66.7&118.1&52.7&78.8&51.8&29.2&88.7&52.8&41&36.6&32.8&59.7&19&22.4&    &32.8&61.7&34.8&68.1&66.2\\
4443.20 &  Fe I  &2.86 &  --1.043  &18.9&41.3&33.7&     &18.7&37.4&19.2&12.3&47&19.9&14.3&13&    &25.2&    &9.1&    &    &15.3&18.2&29.4&     \\
4447.73 &  Fe I  &2.22 &  --1.339  &44.7&74.2&62.4&104.6&45.6&72.6&44&24.5&80.4&42.1&35.6&27.6&23&53.5&17.1&16.8&    &21.2&44.2&33.9&57.8&52.9\\
4466.56 &  Fe I  &2.83 &  --0.600  &49.3&83.6&71.8&    &54.7&84.5&50.7&28.9&90.3&48.8&41&36.2&28.6&63.7&19.3&23.4&    &29.9&    &    &65.1&62.4\\
4489.75 &  Fe I  &0.12 &  --3.899  &40.3&82.1&82.8&136.2&52.8&84.9&46.1&11.6&85.9&44.1&35.1&32.5&5.6&64.3&    &9&    &    &43.6&26.2&56.4&55.3\\
4494.57 &  Fe I  &2.20 &  --1.143  &54.9&85.8&70.2&120.5&60.3&82.9&55.7&34.5&88&51.4&44.5&43.9&41.6&64.9&    &27.4&68.8&31.5&50.3&35.2&65.8&65.9\\
4528.63 &  Fe I  &2.18 &  --0.887  &71&112.9&90.6&153.3&77.4&103.3&73.4&51&114.9&65.9&63.5&55.1&57.2&76.4&32.6&42.8&90.8&40.7&79.5&60.7&89.6&     \\
4531.16 &  Fe I  &1.49 &  --2.101  &48.9&82.4&68&127&55.8&82.6&50.6&21.9&87.6&47.5&39.3&37.4&31.7&    &    &    &70.1&25.3&49.4&38.3&63.1&60.7\\
4602.95 &  Fe I  &1.49 &  --2.208  &46.3&81.2&71.6&121.1&53.1&80.1&48.4&20.2&27.2&44&39.1&32.8&30.8&60.7&22.3&19.8&65.9&    &43.7&32.7&59.8&57\\
4736.78 &  Fe I  &3.21 &  --0.752  &23.1&45.6&32.7&72.6&23.4&43.1&22.5&14.2&53.9&24.4&16.8&16&12.2&24.6&14.4&    &38.8&    &22.5&    &33.2&29\\
4871.32 &  Fe I  &2.87 &  --0.362  &51.2&82.5&64.5&107.9&55.7&79&52.3&36.3&86.5&54.6&44.4&    &37.6&59.5&23.5&33.2&69.2&33.1&54.9&39.2&65.2&63.8\\
4872.14 &  Fe I  &2.88 &  --0.567  &41.5&71.9&55.6&99.8&45.6&67.7&42.8&27.1&75.8&44&32.9&    &29.7&46.3&21.7&27.4&56&21.3&41.3&29.1&53.6&54.3\\
4890.76 &  Fe I  &2.88 &  --0.394  &51&84.5&62.3&109.2&54.9&77.8&50.6&35.7&86.7&50.5&44.3&33.1&32.3&60.4&19.4&30&69.8&    &56.9&37.6&65.2&     \\
4891.50 &  Fe I  &2.85 &  --0.111  &64.5&93.2&79&121.4&67.8&91.6&64&49.5&97.7&66.6&58.7&46.5&48.8&71.4&30.8&43.7&80.5&42.3&61.5&50&76.8&74.1\\
4903.32 &  Fe I  &2.88 &  --0.926  &25.6&55&37.4&80.7&28.4&50.4&26.8&15.9&57.2&27.3&20&17.4&15.3&29.8&    &15&41.6&18&25.2&21.9&40.8&35.3\\
4919.00 &  Fe I  &2.87 &  --0.342  &52.7&83.7&66.9&110.7&56.4&78.9&53&38.2&86.9&54&44.9&39.5&37.1&62.6&28.7&33.2&65.1&30.8&55.8&42.5&68.3&64.2\\
4924.77 &  Fe I  &2.28 &  --2.200  &10.1&29.1&23.6&60.3&14.5&29.7&11.8&4.1&36.2&    &    &7.6&    &    &    &    &19.8&    &10.6&    &19&19.3\\
4938.82 &  Fe I  &2.88 &  --1.077  &18.5&45.6&32.1&72.3&22.4&42.8&21.8&11.7&51.8&22.8&16&9.9&12.3&31.9&5.9&    &    &    &20.1&16.5&35.2&26.4\\
4939.69 &  Fe I  &0.86 &  --3.252  &30&67.8&63.1&117.4&37.9&70.7&33.6&8.6&77.1&32.4&26.6&17.8&20.8&50&    &    &55.9&    &    &21.2&47.1&39.8\\
4946.39 &  Fe I  &3.37 &  --1.010  &7.2&19.6&    &35.3&6.8&16.5&7.4&3.9&21.4&    &    &    &    &    &    &    &    &    &    &    &15.9&    \\ 
4966.09 &  Fe I  &3.33 &  --0.840  &13.9&34.5&21.7&56.2&15.4&31.1&14.5&8.2&39&14&10.8&    &8.7&20.8&    &    &    &    &12&11.2&23.9&17\\
4973.01 &  Fe I  &3.96 &  --0.850  &3.8&9.3&3.2&16.8&4.6&9.7&3.3&2.7&9.8&    &    &    &    &    &    &    &    &    &    &    &    &     \\
4994.14 &  Fe I  &0.92 &  --2.969  &39.8&79.4&70.5&127&50.1&81.9&43&13&83.3&42&31.8&27.8&31&59.9&17.4&    &60.3&18.8&41.4&32.3&50.7&56\\
5001.87 &  Fe I  &3.88 &    0.050  &16.6&35.7&24.2&51.1&15.4&30.3&16.3&13.2&40.3&17.3&6.1&10.8&10.6&23.9&9.1&14.3&29.3&    &18.1&    &28.8&20.2\\
5006.12 &  Fe I  &2.83 &  --0.615  &42.5&73&58.6&101.7&45.2&70&43.1&27.2&78.8&39.3&33.5&28.8&28.4&49.8&15.5&23.4&    &23.6&41.1&27.5&57.6&51.4\\
5014.94 &  Fe I  &3.94 &  --0.270  &10.2&23.7&12.2&37.5&13&22.5&10.1&8&26.7&    &11.1&    &    &13&    &    &    &    &6.4&    &16.2&13.3\\
5022.24 &  Fe I  &3.98 &  --0.490  &6.4&16.3&9.8&26.5&8.5&14.4&7.4&5.6&18.9&    &    &    &    &    &    &    &    &    &    &    &14.5&6.2\\
5041.76 &  Fe I  &1.49 &  --2.203  &52.1&95.3&82.4&141.9&56.1&91.7&53.8&22.1&101.4&53.5&47.1&36.4&33.2&61.7&19.7&22.8&    &25.7&49.8&30.6&67.6&     \\
5044.21 &  Fe I  &2.85 &  --2.040  &3.8&11.4&6.7&23.1&5.1&9.9&3.5&1.7&11.2&    &    &    &    &    &    &    &    &    &    &    &    &     \\
5049.83 &  Fe I  &2.28 &  --1.355  &39.5&73.2&62&109&43&72.3&42.2&21.5&79.4&40.4&32.5&25.3&27&49.8&13.3&20.4&54.9&21.9&38.3&24&56.4&51.8\\
5051.64 &  Fe I  &0.92 &  --2.764  &56.9&96.7&90.5&145.8&67.1&97.6&61.4&23.9&101.8&57&51.7&41.5&44&76.7&21.3&21.7&83&32.2&56.6&41.6&71.4&71.6\\
5068.77 &  Fe I  &2.94 &  --1.041  &17.1&42.7&26.2&68.1&21.7&39.3&19.7&11.7&48.3&    &15.7&17.5&14.2&25.6&    &    &39.1&    &19.2&    &29.1&27.9\\
5074.75 &  Fe I  &4.22 &  --0.160  &8&21.7&12&31.7&9.5&18&8.5&7.9&23.7&    &    &    &    &    &    &    &     &    &7.5&    &17&     \\
5083.35 &  Fe I  &0.96 &  --2.842  &44.1&83.5&76.6&129.7&53.7&84.7&48.3&15.5&89.8&43.3&39.6&29.8&31.3&63.3&14.4&13.2&67.6&    &41.9&28.2&58.5&59.1\\
5090.77 &  Fe I  &4.26 &  --0.360  &4.5&8.5&    &15&4&7.2&3.8&2.4&9.7&    &    &    &    &    &    &    &    &    &    &    &6.6&     \\
5123.73 &  Fe I  &1.01 &  --3.057  &    &    &    &126.7&    &77.5&38.6&    &      &35.7&30.8&22.6&26.6&54.3&    &    &    &    &47.6&    &53.6&     \\
5125.12 &  Fe I  &4.22 &  --0.080  &    &    &    &35.4&    &19.5&8.3&    &      &    &    &    &    &    &    &    &    &    &    &    &16.2&     \\
5127.37 &  Fe I  &0.92 &  --3.249  &    &    &    &117.4&    &70.5&32.8&    &      &    &26.5&19.7&17.6&47.7&    &    &    &    &30.3&    &40.4&     \\
5133.69 &  Fe I  &4.18 &    0.200  &    &    &    &51.5&    &31.8&17.4&    &      &    &12.8&    &    &20.2&    &    &    &    &18.7&    &27.2&19.9\\
5150.85 &  Fe I  &0.99 &  --3.037  &    &    &    &118.3&    &73.5&34.7&    &      &30.7&25.6&22&20.2&50.1&9.7&    &    &    &33.3&    &47.3&45.6\\
5151.92 &  Fe I  &1.01 &  --3.321  &    &    &    &110.5&    &61.5&25.4&    &      &23.3&20.6&15.1&16.8&40.3&    &    &    &    &27.4&    &35.4&30.8\\
5162.28 &  Fe I  &4.18 &    0.020  &    &    &    &43.5&    &27.6&14.8&    &      &    &14.4&11.6&10.1&22.3&    &    &    &    &    &    &26.1&19.7\\
5171.61 &  Fe I  &1.49 &  --1.721  &    &    &    &147.3&    &107.4&75.4&    &      &69.8&68.9&57.2&58.3&86.3&    &    &    &    &69.9&    &80.1&87\\
5191.47 &  Fe I  &3.04 &  --0.551  &    &    &    &100.5&    &60.2&34&    &      &31.2&24.7&21.6&19.7&37.8&10.4&    &    &    &    &    &48.9&     \\
5192.35 &  Fe I  &3.00 &  --0.421  &    &    &    &99.7&    &70.3&41.7&    &      &41.1&34.1&26.2&26&48.1&17&    &    &    &    &    &53.9&51.7\\
5194.95 &  Fe I  &1.56 &  --2.021  &    &    &    &13.1&    &89.6&54.5&    &      &51.6&44.6&39.7&37.8&67.7&17.1&    &    &    &53.3&    &65.2&67.7\\
5198.71 &  Fe I   &2.22 &  --2.140  &    &    &    &130.1&    &38&14.4&    &      &16.7&11.1&11.7&    &    &    &    &    &    &14.4&    &23.7&     \\
4178.86 &  Fe II  &2.58 &  --2.489  &41.2&63.9&47.7&    &35.2&56.1&34&18&56.5&34.3&32.4&20.7&    &43&    &    &    &    &30.2&    &43.5&39.7\\
4233.17 &  Fe II  &2.58 &  --1.900  &    &    &    &    &    &88.3&62.9&43.2&      &    &60.8&50.7&70.4&76.2&    &    &    &    &66.1&    &73.4&75.4\\
4416.83 &  Fe II  &2.78 &  --2.580  &21.2&53.9&30.3&64.8&26.6&41.4&20.8&12&51.6&    &15.8&16.4&    &    &    &13.6&    &    &26.8&16.4&35.4&     \\
4491.41 &  Fe II  &2.85 &  --2.590  &15.8&35.5&24.2&46.6&18.2&30.9&13.8&7.6&36.1&17.2&11.9&    &8.6&26.9&    &4.3&    &    &14.6&14.4&24.2&16.3\\
4508.29 &  Fe II  &2.86 &  --2.318  &26.6&62.9&40.7&71.9&32.1&51.2&29&16.1&57.7&29.9&28.3&20.4&18.4&42.2&12&14.2&50.5&13.2&28.4&22&42.7&37.7\\
4515.34 &  Fe II  &2.84 &  --2.422  &22.3&53.5&34.8&    &25.7&43.9&23&12.6&51.3&25.1&22.1&13.6&16&32.5&    &9.7&37.9&9.5&22.4&15.8&31.7&28.3\\
4520.23 &  Fe II  &2.81 &  --2.590  &21.3&51&28.8&    &25.3&42.8&22.1&13.2&48.5&24.3&19.6&15.9&12.3&27.7&    &9.3&39.7&    &22.2&14.8&27.8&     \\
4522.64 &  Fe II  &2.84 &  --2.050  &    &    &    &    &51.4&65.6&42.7&22.8&      &42.3&38.9&    &    &54.8&18.7&17.9&    &22.3&57.4&34.1&56.3&     \\
4541.52 &  Fe II  &2.86 &  --3.030  &7.7&27.3&18.8&    &11.8&21.9&10.7&    &28.8&    &    &    &    &    &    &    &    &    &17.8&    &14.3&     \\
4555.89 &  Fe II  &2.83 &  --2.304  &26.7&    &39.7&    &31.9&53.1&30.7&    &59.7&    &    &    &    &    &    &15.1&    &    &25.6&    &37.8&35.8\\
4576.34 &  Fe II  &2.84 &  --2.920  &9&28.6&17.4&39.9&12.6&23.3&11.5&    &29.4&    &10.9&    &    &    &    &    &    &    &    &    &18.9&16.2\\
4583.84 &  Fe II  &2.81 &  --1.890  &53.2&92.4&73.8&104.8&58.9&16.9&56.1&    &86.2&52.8&55&40.5&41.9&67.3&22.5&32.6&76.6&29.5&53.9&50.8&66.2&64.6\\
4923.93 &  Fe II  &2.89 &  --1.307  &73.7&114.4&93.7&126.4&82.3&104.2&76.7&56.9&106.7&73.3&79.5&63.2&70&84.8&39.1&49.1&92.2&46.2&76.8&67.9&86&86.9\\
4993.35 &  Fe II  &2.81 &  --3.670  &1.8&8.4&4.7&14.3&4.2&8.8&3.4&1&9.4&    &    &    &    &    &    &    &    &    &    &    &3.1&     \\
5018.45 &  Fe II  &2.89 &  --1.292  &84.8&128.1&106.3&143.6&88.4&115.1&86.1&67.5&117.5&80.2&89.2&71.5&77.3&92.8&47&57.5&106.4&54.5&83.2&79.5&93.9&95.9\\
5197.58 &  Fe II  &3.23 &  --2.167  &    &    &    &54.8&    &39.4&19.3&    &      &    &17.2&12.2&    &28.3&    &    &    &    &    &    &29.3&     \\
3845.47 &  Co I   &0.92 &    0.010 &64.8&91.7&58.7&108.1&69.3&92&67.6&39.6&87.5&66.1&64.4&59.3&48.2&68.3&36.4&37.1&73&39.3&63.6&54.5&70.9&73.4\\
3995.32 &  Co I   &0.92 &  --0.220  &    &    &    &    &61.7&80.4&59.6&32.8&      &54.2&52.2&48.6&49.3&58&32.7&29.1&51.3&37.2&57&    &66.6&68\\
4118.78 &  Co I   &1.05 &  --0.490  &    &    &    &    &    &    &    &18&      &    &    &    &    &    &    &    &    &20.5&    &    &    &     \\
4121.32 &  Co I   &0.92 &  --0.320  &62.9&88.3&72.6&121&65.9&90&64.1&30.4&91.5&61.3&56.6&50.7&54.9&63.4&26.5&27.8&56.8&38.3&  syn  &41.6&71.4&73.3\\
3807.15 &  Ni I   &0.42 &  --1.180  &85.3&111.9&111&145&91.3&116.3&86.3&64.3&113.4&83.3&93.1&71.7&83.1&85.6&52.3&59.1&114.9&70.3&81.5&72.9&90&     \\
3858.30 &  Ni I   &0.42 &  --0.970  &93&123.3&91.4&178.7&97.6&125.6&95.5&72.1&121.4&78.7&97.9&84.3&89.5&87&61.9&63.8&78.5&61.7&90.7&85.3&93.9&106.9\\
5105.55 &  Cu I   &1.39 &  --1.520  &    &    &1.8&19.5&    &3.3&    &    &5.2&    &    &    &    &    &    &    &    &    &    &    &    &     \\
4722.16 &  Zn I   &4.03 &  --0.390  &7.8&21.3&10.5&29.8&8.8&15.1&7.3&3.1&20.6&11.6&7.6&7.3&20.5&10.1&    &    &    &    &14.1&11.2&13.6&8.3\\
4810.54 &  Zn I   &4.08 &  --0.170  &11.6&23.4&14.2&35.6&11.7&20.5&11&5&25.6&8.9&8&11.9&    &14.5&    &    &    &    &9.2&8.5&    &12.2\\
4077.72 &  Sr II  &0.00 &    0.150 &121&  syn  &158.2&  syn  &    &    &    &    &  syn    &    &    &    &    &    &94.4&89.9&    &    &  syn  &100&  syn  &  syn   \\
4161.82 &  Sr II  &2.94 &  --0.600  &    &7.4&    &    &1.6&    &    &    &      &    &    &    &    &    &    &    &134.8&    &    &    &    &     \\
4215.54 &  Sr II  &0.00 &  --0.170  &    &172&147&240.1&129.6&159.6&124.7&66.1&172.6&84.1&46.1&105.1&86&118&81.5&    &93.7&78.8&146.4&73.9&141.3&163.1\\
3774.33 &  Y II   &0.13 &    0.210 &47.7&88.6&    &120.3&64.1&    &    &    &89&    &    &    &    &    &    &20.6&    &    &82.7&    &68.8&89.5\\
3788.70 &  Y II   &0.10 &  --0.070  &40.2&82.7&85.8&111.5&58.3&61.7&47&6.3&79.6&17.7&    &37.4&    &50.8&31.8&19.8&    &35.1&72.3&  &59.4&88.6\\
3818.34 &  Y II   &0.13 &  --0.980  &    &  syn  &    &    &20.5&29.4&14.6&    &      &8&    &8.2&    &18.5&    &    &    &    &32.4&    &30&54.2\\
3950.36 &  Y II   &0.10 &  --0.490  &22.1&67.1&41.3&88.7&39.3&47.4&30.2&4.8&65.1&10.9&    &22.3&    &25.3&8.2&9.4&    &12.8&57.3&    &44.7&71.8\\
4177.54 &  Y II   &0.41 &  --0.160  &    &  syn  &    &    &    &    &    &    &      &    &    &    &    &    &    &    &    &    &81.1&    &    &     \\
4398.01 &  Y II   &0.13 &  --1.000  &10.8&40.9&26.2&    &19.7&30.3&    &    &      &    &    &    &    &    &    &    &    &    &    &    &    &     \\
4883.69 &  Y II   &1.08 &    0.070  &7&38.6&22.3&59.4&16.2&25&11.3&1&41.4&7.3&    &9&    &14.4&    &    &    &    &34.6&    &24.6&41.3\\
5087.43 &  Y II   &1.08 &  --0.170  &4.2&24.4&13.7&44.2&9&13.8&6&    &26.7&    &    &    &    &10.5&    &    &    &    &22.6&    &13.8&32\\
3836.77 &  Zr II  &0.56 &  --0.060  &25.4&52.6&50.6&    &34.7&47.3&    &3.9&57.7&13.2&    &26.2&    &25.6&8.5&    &    &    &50.5&    &45.3&47.1\\
4161.21 &  Zr II  &0.71 &  --0.720  &    &  syn  &    &  syn  &15.9&    &12.2&    &      &5.9&    &    &    &14.5&    &    &    &    &30.3&    &24&43.2\\
4208.99 &  Zr II  &0.71 &  --0.460  &12.6&41.2&27.8&60.7&19.5&27&    &1.4&45.4&7.3&    &12.5&    &12&    &    &    &    &31.3&    &28.3&42.9\\
4317.32 &  Zr II  &0.71 &  --1.380  &    &10.5&7.1&25.6&5.1&5.3&2.1&    &12.7&3.8&    &    &    &    &    &    &    &    &    &    &3.7&14.7\\
3799.35 &  Ru I   &0.00 &  --0.070  &    &    &    &    &    &    &    &    &      &    &    &    &    &    &    &    &    &    &8.5&    &    &     \\
3404.58 &  Pd I   &0.81 &    0.320  &    &  syn  &    &    &  syn  &    &    &    &  syn    &    &    &    &    &    &    &    &    &    &    &    &    &     \\
4554.04 &  Ba II  &0.00 &    0.170  &  syn  &  syn  &  syn  &  syn & syn & syn & syn & syn & syn & syn & syn & syn & syn & syn & syn & syn & syn & syn & syn & syn & syn &  syn   \\
4934.10 &  Ba II  &0.00 &  --0.150  &  syn & syn & syn & syn & syn & syn & syn & syn & syn & syn & syn & syn & syn & syn & syn & syn & syn & syn & syn & syn & syn & syn   \\
3988.52 &  La II  &0.40 &    0.210  &    & syn &    & syn & syn & syn   & syn   &    &  syn    &    &    &    &    &    &    &    &    &    & syn &    & syn & syn \\
3995.75 &  La II  &0.17 &  --0.060  &    & syn &    &  syn  & syn & syn & syn &  &  syn    &    &    &    &    &    &    &    &    &    & syn &    & syn & syn \\
4086.71 &  La II  &0.00 &  --0.070  & syn & syn &    & syn & syn & syn & syn &    &  syn    &    &    &    &    &    &    &    &    &    & syn &    & syn & syn \\
4123.23 &  La II  &0.32 &    0.130  &    & syn   &    & syn & syn &  syn  & syn &    &  syn    &    &    &    &    &    &    &    &    &    &  syn  &    & syn & syn \\
4333.76 &  La II  &0.17 &  --0.060  &    & syn &    &    & syn & syn   &    &    &  syn    &    &    &    &    &    &    &    &    &    &    &    &    & syn \\
5123.01 &  La II  &0.32 &  --0.850  &    &    &    &  syn  &    &    &    &    &      &    &    &    &    &    &    &    &    &    & syn &    &    &  syn   \\
4073.47 &  Ce II  &0.48 &    0.320  &    &15.7&    &12.4&6.2&    &2.7&    &10&    &    &    &    &    &    &    &    &    &    &    &6.4&26.5\\
4083.23 &  Ce II  &0.70 &    0.240  &    &13.6&    &8.1&4&    &1.7&    &7.7&    &    &    &    &    &    &    &    &    &13.8&    &    &20.5\\
4127.38 &  Ce II  &0.68 &    0.240  &    &  syn  &    &    &3.6&    &    &    &9.8&    &    &    &    &    &    &    &    &    &11.5&    &    &20.4\\
4222.60 &  Ce II  &0.12 &  --0.180  &    &17.7&    &18.6&6.1&    &3&    &15.5&    &    &    &    &    &    &    &    &    &22.5&    &11.6&33.4\\
4418.79 &  Ce II  &0.86 &    0.310  &    &6.4&    &4.5&3.4&    &    &    &6.1&    &    &    &    &    &    &    &    &    &12.9&    &10&     \\
4486.91 &  Ce II  &0.30 &  --0.360  &    &10.7&    &12.6&4.5&    &2.1&    &7.9&    &    &    &    &    &    &    &    &    &15.3&    &6&21.5\\
4562.37 &  Ce II  &0.48 &    0.330  &    &19.3&    &17.8&8.1&    &3.1&    &12.5&    &    &    &    &    &    &    &    &    &26.3&    &10.5&35.8\\
4628.16 &  Ce II  &0.52 &    0.260  &    &14.5&    &13.6&4.7&    &    &    &9.9&    &    &    &    &    &    &    &    &    &21.7&    &5.2&30.6\\
3964.81 &  Pr II  &0.05 &    0.090  &    &    &    &    &    &    &    &    &      &    &    &    &    &    &    &    &    &    &20.1&    &    &     \\
4062.80 &  Pr II  &0.42 &    0.660  &    &    &    &12.7&    &  &    &    &      &    &    &    &    &    &    &    &    &    &    &    &    &     \\
4143.14 &  Pr II  &0.37 &    0.370  &    &  syn  &    &    &12&    &6.6&    &      &    &    &    &    &    &    &    &    &    &    &    &14.7&  syn   \\
4408.84 &  Pr II  &0.00 &  --0.010  &    &18.3&    &    &9.3&    &    &    &11.9&    &    &    &    &    &    &    &    &    &    &    &    &     \\
3973.27 &  Nd II  &0.63 &    0.360  &    &19.9&    &    &8.9&    &    &    &14.2&    &    &    &    &    &    &    &    &    &25.6&    &12.1&36.2\\
4018.81 &  Nd II  &0.06 &  --0.850  &    &8.1&    &    & syn   &    &    &    &      &    &    &    &    &    &    &    &    &    &11.2&    &    &13.9\\
4021.34 &  Nd II  &0.32 &  --0.100  &    &17.7&    &14.5&7.8&    &3.4&    &13.9&    &    &    &    &    &    &    &    &    &26.8&    &7.2&35.1\\
4061.09 &  Nd II  &0.47 &    0.550  &    &42.3&    &40.4&20.3&6.1&9.3&1.4&34&    &    &    &    &    &    &    &    &    &51.5&    &20&63.9\\
4069.27 &  Nd II  &0.06 &  --0.570  &    &13.6&    &15.6&    &    &    &    &9.9&    &    &    &    &    &    &    &    &    &22&    &9.1&34.3\\
4109.46 &  Nd II  &0.32 &    0.350  &    &49.9&    &50.8&23.9&    &10.6&    &38.8&    &    &    &    &    &    &    &    &    &61.1&    &27.2&71.4\\
4232.38 &  Nd II  &0.06 &  --0.470  &    &    &    &24.1&    &    &2.8&    &      &    &    &    &    &    &    &    &    &    &24.8&    &10.9&37.4\\
4446.39 &  Nd II  &0.20 &  --0.350  &    &17&    &17.4&7.6&    &2.8&    &14.4&    &    &    &    &    &    &    &    &    &21.8&    &6&35.8\\
4462.99 &  Nd II  &0.56 &  --0.004  &    &17.5&    &19.9&7.1&    &5&    &16.4&    &    &    &    &    &    &    &    &    &27.3&    &9.6&33.9\\
3896.97 &  Sm II  &0.04 &  --0.580  &    &6.6&    &8.5&3.7&    &    &    &5.8&    &    &    &    &    &    &    &    &    &10.5&    &3&23.9\\
4318.94 &  Sm II  &0.28 &  --0.270  &    &14.6&    &18.5&8.6&    &3&    &13.3&    &    &    &    &    &    &    &    &    &21.5&    &6.7&34\\
4519.63 &  Sm II  &0.54 &  --0.430  &    &    &    &8.2&4.1&    &    &    &6.1&    &    &    &    &    &    &    &    &    &11&    &    &20\\
4577.69 &  Sm II  &0.25 &  --0.770  &    &6.4&    &8.7&3.6&    &1.3&    &6.3&    &    &    &    &    &    &    &    &    &11.4&    &    &18.4\\
3819.67 &  Eu II  &0.00 &    0.510  & syn & syn &    & syn & syn &     & syn &    &  syn    &    &    &    &    &    &    &    &    &    & syn   & syn   & syn &     \\
4129.70 &  Eu II  &0.00 &    0.220  &     & syn &    & syn & syn & syn & syn &    &  syn    &    &    &    &    &    &    &    &    &    & syn   & syn   &  syn  &     \\
4205.05 &  Eu II  &0.00 &    0.210  & syn & syn &    & syn & syn &     & syn &    &  syn    &    &    &    &    &    &    &    &    &    & syn   & syn   & syn   &     \\
3768.40 &  Gd II  &0.08 &    0.250  &    &36.9&    &36.4&18.3&4.3&9.3&    &30.9&    &    &    &    &    &    &    &    &    &66&    &25.9&57.9\\
3796.39 &  Gd II  &0.03 &    0.030  &    &    &    &    &27.8&    &    &    &      &    &    &    &    &    &    &    &    &    &  syn  &    &27.8&     \\
3844.58 &  Gd II  &0.14 &  --0.510  &    &15.2&    &    &5.9&    &    &    &10.5&    &    &    &    &    &    &    &    &    &19.5&    &    &32.2\\
3702.86 &  Tb II  &0.13 &    0.440  &    &  syn  &    &9.5&    &    &    &    &      &    &    &    &    &    &    &    &    &    &26.3&    &    &     \\
3848.74 &  Tb II  &0.00 &    0.280  &    &    &    &    &  syn  &    &    &    &      &    &    &    &    &    &    &    &    &    &    &    &    &  syn   \\
4002.57 &  Tb II  &0.64 &    0.100  &    &    &    &  syn  &    &    &    &    &      &    &    &    &    &    &    &    &    &    &6.7&    &    &     \\
4005.47 &  Tb II  &0.13 &  --0.020  &    &    &    &    &    &    &    &    &      &    &    &    &    &    &    &    &    &    &5.1&    &    &  syn   \\
3788.44 &  Dy II  &0.10 &  --0.520  &    &13&    &    &10.8&    &3.3&    &17.2&    &    &    &    &    &    &    &    &    &33.5& &8.1&45.2\\
3869.86 &  Dy II  &0.00 &  --0.940  &    &10.5&    &12.5&5.6&    &    &    &10.1&    &    &    &    &    &    &    &    &    &25.7&    &    &32.9\\
3996.69 &  Dy II  &0.59 &  --0.190  &    &11.7&    &7.5&4.9&    &    &    &10.2&    &    &    &    &    &    &    &    &    &20.4&    &    &29.6\\
4077.96 &  Dy II  &0.10 &  --0.030  &    &  syn  &    &  syn  &  syn  &    &    &    &      &    &    &    &    &    &    &    &    &    &  syn  &    &  syn  &  syn   \\
4103.31 &  Dy II  &0.10 &  --0.370  &    &31.3&    &40.8&18.1&    &5.8&    &24.9&    &    &    &    &    &    &    &    &    &53.1&    &17.3&62.5\\
3692.65 &  Er II  &0.05 &    0.130  &    &48.5&    &38&31.6&    &    &    &41.5&    &    &    &    &    &    &    &    &    &63.6&    &35.3&75.7\\
3786.84 &  Er II  &0.00 &  --0.640  &    &  syn  &    &    &  syn  &    &    &    &  syn    &    &    &    &    &    &    &    &    &    &  syn  &    &    &  syn   \\
3830.48 &  Er II  &0.00 &  --0.360  &    &    &    &    &    &    &    &    &      &    &    &    &    &    &    &    &    &    &43.8&    &14.9&     \\
3896.23 &  Er II  &0.05 &  --0.240  &    &    &    &  syn  &  syn  &    &    &    &  syn    &    &    &    &    &    &    &    &    &    &  syn  &    &27.5&  syn   \\
3938.63 &  Er II  &0.00 &  --0.520  &    &17&    &  syn  &9.9&    &    &    &10.1&    &    &    &    &    &    &    &    &    &29.2&    &7.1&31.2\\
3700.26 &  Tm II  &0.03 &  --0.290  &    &    &    &    &    &    &    &    &      &    &    &    &    &    &    &    &    &    &18.7&    &    &23\\
3701.36 &  Tm II  &0.00 &  --0.420  &    &6.4&    &  syn  &    &    &    &    &4.3&    &    &    &    &    &    &    &    &    &13.9&    &    &13.6\\
3795.76 &  Tm II  &0.03 &  --0.170  &    &7.5&    &5.4&5.6&    &    &    &11.4&    &    &    &    &    &    &    &    &    &22.1&    &8.5&36.3\\
3848.02 &  Tm II  &0.00 &  --0.130  &    &10.9&    &    &  syn  &    &    &    &10.7&    &    &    &    &    &    &    &    &    &22.2&    &9.1&     \\
3694.19 &  Yb II  &0.00 &  --0.299  &    & syn & & syn & syn & & & & syn & & & & & & & & & & syn & & syn & syn    \\
4135.77 &  Os I   &0.52 &  --1.260  &    &    &    &    &    &    &    &    &      &    &    &    &    &    &    &    &    &    &    &    &    &  syn   \\
3513.65 &  Ir I   &0.00 &  --1.260  &    &  syn  &    &    &  syn  &    &    &    &      &    &    &    &    &    &    &    &    &    &    &    &    &     \\
4019.12 &  Th II  &0.00 &  --0.270  &    & syn & & syn & syn & & & & syn & & & & & & & & & & syn & & syn & syn   \\

\tableline
\enddata
~ \\
$^{\rm a}$ Identification of objects is given in Table~\ref{tab:obs}.\\
``syn'' denote that the abundance derived by spectral synthesis
 technique.
\end{deluxetable}

\clearpage

\begin{deluxetable}{clclcl} 
\tablewidth{0pt}
\tablecaption{RADIAL VELOCITIES COMPARED WITH PREVIOUS STUDIES\label{tab:obs3}}
\startdata
\tableline
\tableline
\multicolumn{1}{c}{No} & Object & Radial velocity & References\\
\tableline
1&HD~4306 &  --69.69$\pm$0.29 & This work (19 Aug, 2000) \\
1&HD~4306 & --63.3$\pm$0.2 & McWilliam et al. 1995\\
\tableline\\
2&HD~6268 &  39.20$\pm$0.27  & This work (18 Aug, 2000) \\
2&HD~6268 &  38.9$\pm$0.3 & McWilliam et al. 1995\\
\tableline\\
6&HD~122563 & --27.20$\pm$0.33 & This work (4 July, 2000)\\
6&HD~122563 & --26.52$\pm$0.34 & This work (29 Jan, 2001)\\
6&HD~122563 & --27.0$\pm$0.34  & Norris et al. 1996\\
\tableline\\
7&HD~126587 &  148.72$\pm$0.72 & This work (27 Jan, 2001)\\
7&HD~126587 &   149.10$\pm$0.24 & This work (31 Jan, 2001)\\
7&HD~126587 & 150.0$\pm$0.3  & McWilliam et al. 1995\\
7&HD~126587 & 149.40$\pm$0.17& Carney et al. 2003\\
\tableline\\
8&HD~140283 &   --171.17$\pm$0.29 & This work (4 July, 2000)\\
8&HD~140283 & --170.23$\pm$0.19 & This work (17 Aug, 2000)\\
8&HD~140283 & --171.9$\pm$0.19 & Norris et al. 1996\\
\tableline\\
9&HD~186478 &  30.52$\pm$0.31 & This work (20 Aug, 2000)\\
9&HD~186478 & 32.7$\pm$0.3 & McWilliam et al. 1995\\
9&HD~186478 & 30.43$\pm$0.12* & Carney et al. 2003\\
\tableline\\
19&CS~22892--052 &   12.72$\pm$0.49 & This work (22 July, 2001)\\
19&CS~22892--052 & 13.1$\pm$0.3 & McWilliam et al. 1995\\
19&CS~22892--052 & 13.6$\pm$0.3 & Norris et al. 1996\\
19&CS~22892--052 & 12.48$\pm$0.61* & Preston \& Sneden 2001\\
\tableline\\
20&CS~22952--015 &  --20.07$\pm$0.71 & This work (11 Nov, 2000)\\
20&CS~22952--015 & --18.8$\pm$0.4 & McWilliam et al. 1995\\
20&CS~22952--015 & --19.2$\pm$1 & Norris et al. 1996\\
\tableline\\
22&CS~31082--001 &   138.91$\pm$0.30 & This work (30 July, 2001)\\
22&CS~31082--001 & 139.05$\pm$0.05 & Hill et al. 2001\\
\tableline
\enddata
~ \\

* Mean value of their measurements
\end{deluxetable}

\clearpage

\end{document}